\documentclass[a4paper,11pt]{article}
\pdfoutput=1 

\usepackage{jheppub} 

\usepackage[T1]{fontenc} 
\usepackage[textsize=footnotesize,textwidth=2.5cm]{todo notes}

\definecolor{mikadoyellow}{rgb} {0.16, 0.254, 0.6}

\usepackage{amssymb,amsmath,latexsym,bm,amsfonts}
\usepackage{graphicx}
\usepackage{longtable}
\usepackage{color,xcolor}
\usepackage{indentfirst}
\usepackage{subfigure}
 \usepackage{comment}

\def\p{\partial}

\title{ \boldmath Towards the generalized gravitational entropy for spacetimes with  non-Lorentz invariant duals }

\author[a,b]{Qiang Wen}

\affiliation[a]{Shing-Tung Yau Center and School of Mathematics, Southeast University, Nanjing 210096, China}
\affiliation[b]{Graduate School, China Academy of Engineering Physics, Beijing 100193, China}

\emailAdd{101012491@seu.edu.cn}

\abstract{Based on the Lewkowycz-Maldacena prescription and the fine structure analysis of holographic entanglement proposed in \cite{Wen:2018whg}, we explicitly calculate the holographic entanglement entropy for warped CFT that duals to AdS$_3$ with a Dirichlet-Neumann type of boundary conditions. We find that certain type of null geodesics emanating from the entangling surface $\partial\mathcal{A}$ relate the field theory UV cutoff and the gravity IR cutoff. Inspired by the construction, we furthermore propose an intrinsic prescription to calculate the generalized gravitational entropy for general spacetimes with non-Lorentz invariant duals. Compared with the RT formula, there are two main differences. Firstly, instead of requiring that the bulk extremal surface $\mathcal{E}$ should be anchored on $\partial\mathcal{A}$, we require the consistency between the boundary and bulk causal structures to determine the corresponding $\mathcal{E}$. Secondly we use the null geodesics (or hypersurfaces) emanating from $\partial\mathcal{A}$ and normal to $\mathcal{E}$ to regulate $\mathcal{E}$ in the bulk. We apply this prescription to flat space in three dimensions and get the entanglement entropies straightforwardly.}

\begin{document} 
\maketitle
\flushbottom
\section{Introduction}
Entanglement entropy, which describes the correlation structure of a quantum system, has played a central role in the study of modern theoretical physics. In the context of AdS/CFT correspondence \cite{Maldacena:1997re,Gubser:1998bc,Witten:1998qj}, the Ryu-Takayanagi (RT) \cite{Ryu:2006bv,Ryu:2006ef} formula relates the entanglement entropy to a geometric quantity on the gravity side.  More explicitly, for a static subregion $\mathcal{A}$ in the boundary CFT and a minimal surface $\mathcal{E}_{\mathcal{A}}$ in the dual AdS bulk that anchored on the boundary $\partial\mathcal{A}$ of $\mathcal{A}$, the RT formula states that the entanglement entropy of $\mathcal{A}$ is measured by the area of $\mathcal{E}_{\mathcal{A}}$ in Planck units,
\begin{align}\label{RT}
S_{\mathcal{A}}=\frac{Area (\mathcal{E}_{\mathcal{A}})}{4 G}\,.
\end{align}
Soon the Hubeny-Rangamani-Takayanagi (HRT) \cite{Hubeny:2007xt} formula was proposed as the covariant version of the RT formula. Accordingly the minimal surface is generalized to the extremal surface in the HRT formula. The holographic picture of entanglement entropy has a huge impact on our understanding of holography itself as well as the emergence of spacetime.
 
One way to understand the RT formula is the Rindler method, which is first proposed in \cite{Casini:2011kv} and later generalized in \cite{Song:2016gtd,Jiang:2017ecm}. The key point of the Rindler method is to construct a Rindler transformation, which is a symmetry of the theory, that maps the causal development of a subregion to a thermal ``Rindler space''. So the problem of calculating the entanglement entropy of a subregion is replaced by the problem of calculating the thermal entropy of the Rindler space. According to holography, the thermal entropy of the Rindler space equals to the thermal entropy of its bulk dual, which is usually a hyperbolic black hole (or black string). The horizon of the hyperbolic black hole is exactly what maps to the RT surface under the Rindler transformations in the bulk.

The other way is to extend the replica trick into the bulk and calculate the entanglement entropy using the partition function calculated by the path integral on the gravity side. This prescription is explicitly carried out by Lewkowycz and Maldacena \cite{Lewkowycz:2013nqa} (see \cite{Dong:2016hjy} for the covariant generalization). The entanglement entropy of a quantum system is defined as the von Neumann entropy $S_\mathcal{A}=-\text{Tr}\rho_{\mathcal{A}}\log\rho_{\mathcal{A}}$ of the reduced density matrix $\rho_\mathcal{A}$. Consider a quantum field theory on $\mathcal{B}$, the replica trick first calculates the Renyi entropy $S^{(n)}_{\mathcal{A}}=\frac{1}{1-n}\log\text{Tr}\rho_{\mathcal{A}}^{n}$ for $n=\mathbb{Z}_{+}$, then analytically continues $n$ away from integers. We get the entanglement entropy $S_{\mathcal{A}}$ when $n\to 1$. To calculate Tr$\rho^{n}_{\mathcal{A}}$, we can cut $\mathcal{B}$ open along $\mathcal{A}$, glue n copies of them cyclically into a new manifold $\mathcal{B}_n$ and then do path integral on $\mathcal{B}_n$. The entanglement entropy is calculated by
\begin{align}
S_{\mathcal{A}}=-n\partial_n \left(\log \mathcal{Z}_n-n\log\mathcal{Z}_1\right)|_{n=1}\,,
\end{align}
where $\mathcal{Z}_n$ is the partition function of the quantum field theory on $\mathcal{B}_n$. Assuming holography and the unbroken replica symmetry in the bulk, the LM (Lewkowycz-Maldacena) prescription manages to construct the bulk dual of $\mathcal{B}_n$, which is a replicated bulk geometry $\mathcal{M}_n$ with its boundary being $\mathcal{B}_n$. Then the partition function $\mathcal{Z}_n$ can be calculated by the path integral on $\mathcal{M}_n$ on the gravity side. The two main results of \cite{Lewkowycz:2013nqa,Dong:2016hjy} are:
\begin{itemize}
\item the holographic entanglement entropy is calculated by the area of the codimension two surface $\mathcal{E}$ in Plank units, which is the set of all the fixed points of the bulk replica symmetry, 
\item the codimension two surface $\mathcal{E}$ is an extremal surface{\footnote{ The extremal condition is the result of imposing the equations of motion and replica symmetry on all the fields in the action.  In \cite{Song:2016pwx}, as the gauge fields are nondynamical and do not appear in the symplectic structure, thus should not be imposed with the replica symmetry (or periodic) condition. As a result, in that case the geometric quantity $\mathcal{E}$ that measures the entanglement entropy is not an extremal surface. See \cite{Dong:2017xht} for a simpler discussion on the extremal condition.}}.
\end{itemize}  
As was indicated in \cite{Lewkowycz:2013nqa,Dong:2016hjy}, these results are quite general even for holographies beyond AdS/CFT.

However the above results are not equivalent to the RT (or HRT) formula without the homology constraint and the prescription to regulate the entanglement entropy via the UV/IR cutoff relation \cite{Susskind:1998dq} in AdS/CFT. The homology constraint requires the extremal surface $\mathcal{E}$ to be anchored on $\partial \mathcal{A}$ and homologous to $\mathcal{A}$ \cite{Headrick:2007km,Headrick:2013zda,Headrick:2014cta}, thus selects the right extremal surface that matches $\mathcal{A}$. The prescription for regulation tells us how to regulate the extremal surface $\mathcal{E}$ in the bulk when we regulate $\mathcal{A}$ on the boundary. Although the homology constraint and prescription for regulation in the RT formula seems quite natural, it has never been thoroughly studied in holographies beyond AdS/CFT.

In the context of AdS/CFT, since both of the boundary field theory and the bulk gravity are relativistic, the causal structures near the entangling surface $\partial\mathcal{A}$ and the RT surface $\mathcal{E}$ are consistent with each other. This naturally leads to the requirement that $\mathcal{E}$ should be anchored on $\partial\mathcal{A}$\footnote{A proof for the homology constraint at topological level in AdS/CFT is given in \cite{Haehl:2014zoa}}. However, for holographies beyond AdS/CFT{\footnote{Although the AdS/CFT has attracted most of the attentions, the holographic principle is assumed to be hold for general spacetimes. So far the holography beyond AdS/CFT that has been proposed include the dS/CFT correspondence \cite{Strominger:2001pn}, the Lifshitz spacetime/Lifshitz-type field theory duality \cite{Son:2008ye,Balasubramanian:2008dm,Kachru:2008yh,Taylor:2015glc}, the Kerr/CFT correspondence \cite{Guica:2008mu}, the WAdS/CFT \cite{Anninos:2008fx,Compere:2014bia} or WAdS/WCFT \cite{Detournay:2012pc,Song:2017czq} correspondence, and flat holography in four dimensions \cite{Bondi:1962px,Sachs:1962wk,Strominger:2013jfa} and three dimensions \cite{Barnich:2010eb,Bagchi:2010eg,Bagchi:2012cy}.}}, especially those with non-Lorentz invariant field theory duals (for example non-relativistic theories, ultra-relativistic theories or Lifshitz-type theories), it is reasonable to question the validity of the homology constraint. 
Also the UV/IR cutoff relations and their application to regulate the bulk extremal surface in more general holographies have not been discussed before. These are crucial to the validity of the RT formula.

Recently a series of work \cite{Song:2016gtd,Song:2016pwx,Jiang:2017ecm} calculated the holographic entanglement entropy for spacetimes that are not asymptotically AdS and found the corresponding geometric quantities. Remarkably, these results challenge the validity of the RT formula. In the context of (warped) AdS/warped CFT correspondence \cite{Detournay:2012pc,Compere:2013bya} and 3-dimensional flat holography \cite{Barnich:2010eb,Bagchi:2010eg,Bagchi:2012cy}, the geometric quantities $\mathcal{E}_{\mathcal{A}}$, which calculate the entanglement entropy of a single interval $\mathcal{A}$ in warped CFT (WCFT) \cite{Detournay:2012pc} and BMS$_3$ invariant field theories (BMSFTs), are found \cite{Song:2016gtd,Jiang:2017ecm} respectively with the Rindler method. In both cases, the holographic calculations consist with the field theory results \cite{Castro:2015csg,Azeyanagi:2018har,Bagchi:2014iea,Basu:2015evh}. The corresponding geometric quantities that satisfy \eqref{RT} are spacelike geodesics in the bulk, thus consistent with the results in \cite{Lewkowycz:2013nqa,Dong:2016hjy}. However, unlike the RT surfaces, the endpoints of $\mathcal{E}_{\mathcal{A}}$ are not anchored on $\partial\mathcal{A}$. 

For example, in 3-dimensional flat space, the endpoints of $\mathcal{E}_{\mathcal{A}}$ are in the bulk and connected to $\partial\mathcal{A}_{\pm}$ by two null geodesics $\gamma_\pm$ normal to $\mathcal{E}_{\mathcal{A}}$ \cite{Jiang:2017ecm}. Recent works related to this geometric picture can be found in \cite{Hijano:2017eii,Asadi:2018lzr,Fareghbal:2018ngr}. This new geometric picture of entanglement entropy with the extra null geodesics $\gamma_\pm$ is reformulated in \cite{Hijano:2017eii} following the HRT covariant formulation \cite{Hubeny:2007xt}.

\begin{figure}[h] 
   \centering
        \includegraphics[width=0.7\textwidth]{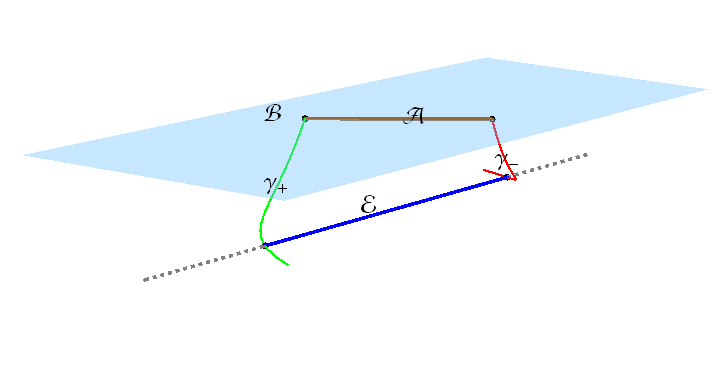}    
 \caption{ The blue solid line is the $\mathcal{E}_{\mathcal{A}}$ which is regulated from the spacelike geodesic $\mathcal{E}$. The red and green lines are the null geodesics $\gamma_{\pm}$ that connect the endpoints of $\mathcal{E}_{\mathcal{A}}$ and $\mathcal{A}$.
\label{geometricfigure} }
\end{figure}

For (warped) AdS$_3$ which duals to a WCFT, the geometric picture for holographic entanglement entropy is also a spcaelike geodesic $\mathcal{E}_{\mathcal{A}}$ with endpoints in the bulk \cite{Song:2016gtd}. We will show that (which is not adressed in \cite{Song:2016gtd}) the endpoints of $\mathcal{E}_{\mathcal{A}}$ are also connected to the endpoints of $\mathcal{A}$ at the cutoff boundary by two null geodesics $\gamma_\pm$, which are normal to $\mathcal{E}_{\mathcal{A}}$ (see Fig.\ref{geometricfigure}). 

The above results also imply that the prescription to regulate the $\mathcal{E}$ via the UV/IR cutoff relations is different from the RT formula. Instead of being cut off at an infinitely large radius, the IR cutoff of the curve $\mathcal{E}$ in these cases are at finite radius and in some way controlled by the null geodesics $\gamma_\pm$ emanating from $\partial\mathcal{A}$. In this paper we try to understand this new geometric picture following the LM prescription \cite{Lewkowycz:2013nqa}. We focus on the case of AdS$_{3}$/WCFT correspondence. In this holography the gravity side is AdS$_3$ with the Compere-Song-Strominger (CSS) \cite{Compere:2013bya} boundary conditions while the field theory side is a WCFT. We study the replica story both on the boundary and in the bulk and try to understand the role of the null geodesics $\gamma_\pm$ in the replica story.

We will not re-derive the two main results of \cite{Lewkowycz:2013nqa,Dong:2016hjy} listed above and admit they are true in general holographies. However, to intrinsically determine the geometric picture for holographic entanglement entropy we need to solve the remaining two problems:
\begin{itemize}
\item how to determine the bulk extremal surface $\mathcal{E}$ in the bulk that matches $\mathcal{A}$ if we do not require $\mathcal{E}$ to be anchored on $\partial\mathcal{A}$?

\item how to regulate $\mathcal{E}$ in the bulk accordingly when we regulate $\mathcal{A}$ on the boundary?
\end{itemize}
The above two problems are the main tasks of this paper. To solve the first problem, we need to explicitly study the causal structure of the boundary field theory and find its match in the bulk. In this paper we will not discuss topological ambiguities to determine $\mathcal{E}$. To solve the second problem, we use the prescription of \cite{Wen:2018whg} to study the fine structure in holographic entanglement. More explicitly we use a new geometric quantity named the modular plane, to slice the entanglement wedge. Under this construction we get a fine correspondence between the points on $\mathcal{A}$ and the points on $\mathcal{E}$. We will show that the point where we cut $\mathcal{A}$ off and the point where we cut $\mathcal{E}$ off are just related by this fine correspondence.

The structure of this paper is in the following. In section \ref{secrindlermethod} we present some interesting observations from the Rindler method which partially inspire the intrinsic prescription we will propose. Then we focus on the case of AdS$_3$/warped CFT correspondence. In section \ref{3} we apply the Rindler method to this case. In section \ref{secmodularflows} we will explicitly study the bulk and boundary modular flows and define the modular planes. In section \ref{secgeneralizedge}, we calculate the generalized gravitational entropy for AdS$_3$ with CSS boundary conditions with the help of modular planes. The goal of this section is to understand how do the null geodesics (or modular planes) relate the boundary and bulk cutoffs. Based on the above construction, in section \ref{discussion} we propose an intrinsic prescription of calculate the generalized gravitational entropy for spacetimes with non-Lorentzian duals. At last, we apply our prescription to 3-dimensional flat holography in section \ref{flat}. In appendix \ref{geodesics}, we classify the spacelike geodesics in AdS$_3$. The appendix \ref{appendixA} and \ref{appendixB} are written for special readers, who are interested in the saddle point condition for $\mathcal{E}_{\mathcal{A}}$ and the entanglement contour for WCFT.

\section{New observations from the Rindler method}\label{secrindlermethod}
\subsection{A brief introduction to Rindler method}\label{introductionRM}
In the field theory, the key step of the Rindler method is to construct a Rindler transformation, a symmetry transformation which maps the calculation of entanglement entropy to thermal entropy. The general strategy to construct Rindler transformations and their bulk extensions by using the symmetries of the field theory and holographic dictionary, is summarized in the section 2 of \cite{Jiang:2017ecm}. Here we just give the main points of the Rindler method.

Consider a QFT on manifold $\mathcal{B}$ with the symmetry group $G$. The global symmetries, whose generators are denoted by ${h_j}$, form a subset of $G$. The Rindler transformations $R$, which map a subregion $\mathcal{D}$ of $\mathcal{B}$ to a Rindler space $\tilde{\mathcal{B}}$ with infinitely far away boundary, can be constructed by imposing the following requirements:
\begin{enumerate}
\item The transformations $R:~\tilde{x}^i=f(x^i)$ should be a symmetry transformation of the QFT. 

\item 
The vectors $\p_{\tilde{x}^i}$ in the Rindler space should be a linear combination of the global generators in the original space
\begin{align}\label{vacuum}
\partial_{\tilde{x}^i}=\sum_j b_{ij} h_j\,,
\end{align}
where $b_{ij}$ are arbitrary constants.

\end{enumerate}
The first requirement will give constraints on the coefficients $b_{ij}$. The remaining independent coefficients will control the size, position of $\mathcal{D}$ and the thermal circle of $\tilde{\mathcal{B}}$. Note that the shape of $\mathcal{D}$ is determined by the symmetries and independent of the choice of $b_{ij}$.  The Rindler transformation $R$ will be invariant under some imaginary identification of the Rindler coordinates $\tilde{x}^i \sim \tilde{x}^i+i {\tilde{\beta}}^i$.  This identification can be referred to as a ``thermal'' identification in $\tilde{\mathcal{B}}$.

Our strategy to construct Rindler transformations only involves the global generators, thus has a natural  extension in the bulk. According to the holographic dictionary, the global generators $h_i$ of the asymptotic symmetry group are dual to the isometries of the bulk sapcetime. Then by replacing the $h_i$ generators with the generators of the bulk isometries and requiring the Rindler bulk space to satisfy the same boundary conditions, we can get the Rindler transformations in the bulk. The bulk extension of $\tilde{\mathcal{B}}$ (or the Rindler bulk) should have a horizon whose Bekenstein-Hawking entropy gives the thermal entropy of the field theory on $\tilde{\mathcal{B}}$. Using the inverse bulk Rindler transformations we can map this horizon back to the bulk extension of the original field theory on $\mathcal{B}$. The image of this mapping will give the corresponding geometric quantity $\mathcal{E}$. One can consult \cite{Song:2016gtd,Jiang:2017ecm} for explicit examples of the Rindler method.

\subsection{New observations from the Rindler method}
In \cite{Jiang:2017ecm}, with the inverse bulk Rindler transformations, the authors made several interesting observations.
\begin{itemize}
\item \textit{Observation 1}: the horizon of the Rindler bulk space is mapped to two codimension one null hypersurfaces $\mathcal{N}_{\pm}$ in the original bulk space. The curve $\mathcal{E}$ that related to entanglement entropy is the curve where $\mathcal{N}_-$ intersect with $\mathcal{N}_+$,
\begin{align}\label{way1}
\mathcal{E}=\mathcal{N}_-\cap\mathcal{N}_+\,.
\end{align}

\item \textit{Observation 2}: the Hamiltonian in the Rindler bulk space and boundary $\tilde{\mathcal{B}}$ will be mapped to the modular Hamiltonian in the original bulk and boundary respectively. A modular flow $k_t$ ($k_t^{bulk}$) is generated by the modular Hamiltonian. Then the curve $\mathcal{E}$ is determined by
\begin{align}\label{way2}
k_t^{bulk}(\mathcal{E})=0\,,
 \end{align}
which means $\mathcal{E}$ is the fixed points of the modular flow. 

\item \textit{Observation 3}: the two null hypersurfaces $\mathcal{N}_\pm$ are normal to $\mathcal{E}$. Furthermore their intersection with the boundary $\mathcal{B}$ gives a decomposition on $\mathcal{B}$, which is consistent with the causal structure of the dual field theory. 
\end{itemize}
 
Although these observations are made in 3d flat holography, the logic behind them should work for general holographies. The first observation is just a mapping from the Rindler bulk to the original spacetime and follows the logic of Rindler method. The second way is equivalent to the statement that $\mathcal{E}$ is the fixed points of the bulk replica symmetry which is a key statement of the LM prescription \cite{Lewkowycz:2013nqa}. In the third observation, the requirement that the bulk and boundary causal structure should be consistent is obviously a requirement of holography\footnote{In the context of AdS/CFT this requirement has been discussed in detail in \cite{Headrick:2014cta}.}.  Later we will explicitly show that the above observations are also true in the context of AdS$_3$/WCFT.

Note that, the ways \eqref{way1} and \eqref{way2} to determine $\mathcal{E}$ are not intrinsic and rely on the explicit information of the Rindler transformations and locally defined modular Hamiltonian, which may not even exist. However the third observation implies that, on the other way around, $\mathcal{E}$ can be determined by the requirement that, the bulk causal decomposition by the normal null hypersurfaces $\mathcal{N}_{\pm}$ of $\mathcal{E}$ should reproduce the causal structure of the boundary field theory associated to $\mathcal{A}$, i.e.
\begin{align}\label{causalconsistence}
\mathcal{E}:~~\mathcal{N}_{\pm}\cap\mathcal{B} \supset \partial\mathcal{D}_{\mathcal{A}}
\end{align} 
This prescription to determine $\mathcal{E}$ is intrinsic and can be applied for general spacetimes, thus finishes our first task.

The above construction remind us of the construction of the light-sheet by Bousso \cite{Bousso:2002ju}. In \cite{Hubeny:2007xt}, using the light-sheets the authors propose a prescription to construct the covariant geometric picture for entanglement entropy in the context of AdS/CFT. They require the light-sheet associated to $\mathcal{E}$ should intersect with the boundary on the boundary light-sheet associated to $\partial\mathcal{A}$. This is equivalent to our requirement of the consistency between the bulk and boundary causal structures \eqref{causalconsistence}.

\section{Rindler method applied on the AdS$_3$/WCFT correspondence}\label{3}

\subsection{AdS$_3$ with CSS boundary conditions}\label{2.2}
Let us give a quick review on the AdS$_3$/WCFT correspondence. In the Fefferman-Graham gauge, solutions to 3-dimensional Einstein gravity with a negative cosmological are in the following, 
\begin{align}
{ds^2\over\ell^2}={d\eta^2\over \eta^2}+\eta^2\Big(g^{(0)}_{ab}+{1\over \eta^2}g^{(2)}_{ab}+{1\over \eta^4}g^{(4)}_{ab}\Big)dx^a dx^b\,,
\end{align}
where $\eta$ is the radial direction, and $x^a,\, a=1,2$ parametrize the boundary. Under the Dirichlet boundary (Brown-Henneaux) conditions $\delta g_{ab}^{(0)}=0$, the asymptotic symmetry are generated by two copies of Virasoro algebra \cite{Brown:1986nw}, which indicates the dual field theory is a CFT$_2$. 

In \cite{Compere:2013bya}, a Dirichlet-Neumann type of boundary conditions
\begin{align}\label{css}
\delta g^{(0)}_{\pm-}=0\,,\quad    \p_- g^{(0)}_{++}=0\,,\quad
\delta g^{(2)}_{--}=0\,,
\end{align}
is considered for AdS$_3$ in Einstein gravity, which we refer as the CSS boundary conditions. Under the CSS boundary conditions the metric on the boundary is no longer fixed and is allowed to fluctuate. Without a fixed background metric the usual way to determine the causal structure of the boundary field theory with null geodesics (or hypersurfaces) associated to $\partial\mathcal{A}$ is meaningless. In these cases, one can still define the causal development $\mathcal{D}_\mathcal{A}$ as the subregion in $\mathcal{B}$, which is mapped to the whole Rindler space $\tilde{\mathcal{B}}$ under Rindler transformations  (see section \ref{introductionRM}). This definition of causal development is more general, and will reduce to the definition using null lines when the boundary has a fixed background metric.

Consider a BTZ metric 
\begin{align}\label{btz}
ds^2=\ell^2 \left(T_u^2 du^2+2 r du dv+T_v^2 dv^2+\frac{dr^2}{4 \left(r^2-T_u^2 T_v^2\right)}\right)\,.
\end{align}
When we impose the CSS boundary conditions, the asymptotic symmetry group is featured by a Virasoro-Kac-Moody algebra \cite{Compere:2013bya}, 
\begin{align}\label{tildealgebra}
[\tilde{L}_n,\tilde{L}_m]=&(n-m)\tilde{L}_{n+m}+\frac{\tilde{c}}{12}(n^3-n)\delta_{n+m}\,,
\cr
[\tilde{L}_n,\tilde{P}_m]=&-m\tilde{P}_{m+n}+m\tilde{P}_0\delta_{n+m}\,,
\cr
[\tilde{P}_n,\tilde{P}_m]=&\frac{\tilde{k}}{2}n\delta_{m+n}\,,
\end{align}
with the central charge and Kac-Moody level given by 
\begin{align}
\tilde{c}=&\frac{3\ell}{2G}\,, \qquad \tilde{k}=-\frac{\ell T_v^2}{ G}\,.
\end{align}
Here we have set $T_v$ fixed to satisfy the boundary condition $\delta g^{(2)}_{--}=0$. Then $u$ is the direction that keeps the $SL(2,R)$ global symmetry. Although the local isometries in the bulk are still $SL(2,R)\times SL(2,R)$, only the $SL(2,R)\times U(1)$ part consists with the boundary conditions. 

The above asymptotic symmetry analysis indicates that the dual field theory is a WCFT featured by the algebra \eqref{tildealgebra}. The WCFT has the following local symmetries \cite{Detournay:2012pc}
\begin{align}\label{localsymmetry}
u=f(u')\,,\qquad v=v'+g(u')\,.
\end{align}
A Similar story of asymptotic symmetry analysis happens for warped AdS$_3$ \cite{Compere:2009zj}, and the WAdS/WCFT correspondence \cite{Detournay:2012pc} is conjectured.

The (warped) AdS$_3$/WCFT correspondence has passed several key tests by matching the thermal entropy \cite{Detournay:2012pc}, entanglement entropy \cite{Castro:2015csg,Song:2016gtd}, correlation functions \cite{Song:2017czq} and one-loop determinants \cite{Castro:2017hpx}. See \cite{Compere:2013aya,Hofman:2014loa,Jensen:2017tnb} for a few examples of WCFT models.

Note that the Kac-Moody level in \eqref{tildealgebra} is charge dependent. This is different from the canonical warped CFT algebra \cite{Detournay:2012pc} which has a constant Kac-Moody level. These two algebras can be related by a state dependent coordinate transformation. One can also obtain the canonical algebra using the state-dependent asymptotic Killing vectors \cite{Apolo:2018eky}. The mapping between the entanglement entropies of the theories featured by these two algebras is explicitly discussed in \cite{Song:2016gtd}. In this paper, by WCFT we mean the theory featured by the algebra \eqref{tildealgebra}, and will not do the further mapping to the canonical ones.




\subsection{Rindler method for AdS$_3$ with CSS boundary conditions}\label{Rindlercss}
The global symmetries of the asymptotic symmetry group of AdS$_3$ with CSS boundary conditions are $SL(2,R)\times U(1)$, which consist of the following generators
\begin{align}
 J_-=  \partial _u, \quad J_0=  u \p_{u}-r  \p_{r }, \quad J_+=  u^2\p_{u}-\frac{1}{2 r }\p_{v}-2 r  u \p_{r },
\quad
\bar{J}=  \partial _v\,,
 \end{align} 
Now we consider the AdS$_3$ with $T_u=0,T_v=1$, and the AdS radius being $\ell=1$,
\begin{align}\label{tu0tv1}
ds^2=2 r du dv+(dv)^2+\frac{ (dr)^2}{4 r^2}\,.
\end{align}
On the boundary we choose the interval to be
\begin{align}\label{interval}
\mathcal{A}:~~\{(-\frac{l_u}{2},-\frac{l_v}{2})\to(\frac{l_u}{2},\frac{l_v}{2})\}\,,
\end{align}
One can consult \cite{Song:2016gtd} for the cases with general temperatures.

Following the strategy presented in section \ref{introductionRM}, we can construct the most general Rindler transformations (see appendix A in \cite{Song:2016gtd}). The coefficients $b_{ij}$ control the position, the size of $\mathcal{D}$ and the thermal circle of $\tilde{\mathcal{B}}$. Here for simplicity, we settle down the position of $\mathcal{D}$ and the thermal circle of $\tilde{\mathcal{B}}$, which do not affect the entanglement entropy. Then we get the Rindler transformations from the AdS$_3$ \eqref{tu0tv1} 
to a Rindler $\widetilde{\text{AdS}_{3}}$ with $T_{\tilde{u}}=T_{\tilde{v}}=1$
\begin{align}
ds^2=d\tilde{u}^2 +2 \tilde{r} d\tilde{u} d\tilde{v}+d\tilde{v}^2+\frac{d\tilde{r}^2 }{4 \left(\tilde{r}^2-1\right)}\,,
\end{align}
The Rindler transformations are given by
\begin{align}\label{tutvtr}
\tilde{u}=&\frac{1}{4} \log \left(\frac{R_-^2-1}{R_+^2-1}\right)\,,
\cr
\tilde{v}=&\frac{1}{4} \log \left(\frac{(R_+-1) (R_-+1)}{(R_++1) (R_--1)}\right)+v\,,
\cr
\tilde{r}=&\frac{(R_+-1) (R_--1)}{R_++R_-}+1\,,
\end{align}
where
\begin{align}
R_{\pm}=r (l_u \mp 2 u)\,.
\end{align}

Asymptotically, we get
\begin{align}\label{boundarymap}
\tilde{u}=\text{Arctanh}\left(\frac{2 u}{l_u}\right)\,,
\qquad
\tilde{v}=v\,,
\end{align}
which, as expected, is a warped conformal mapping \eqref{localsymmetry}.
We see that the $(\tilde{u},\tilde{v})$ coordinates only cover a strip-like subregion 
\begin{align}\label{mathcalD}
\mathcal{D}:~-\frac{l_u}{2}<u<\frac{l_u}{2}\,.
\end{align}
We define this strip $\mathcal{D}$ as the causal development of the interval $\mathcal{A}$ (see also \cite{Castro:2015csg}) in WCFT.

The bulk Rindler transformations \eqref{tutvtr} map the horizon of the Rindler $\widetilde{\text{AdS}_{3}}$ at $\tilde{r}=1$ to two null hypersurfaces $\mathcal{N}_{\pm}$,
\begin{align}\label{lightsheetwads}
\mathcal{N}_+:~~ r= \frac{1}{l_u-2 u},\qquad \mathcal{N}_-:~~r= \frac{1}{l_u+2 u}.
\end{align}
We find $\mathcal{N}_{\pm}$ intersect at a curve in the bulk
\begin{align}\label{gammawads}
\mathcal{E}=\mathcal{N}_{-}\cup\mathcal{N}_{+}:~~~~\left\{u=0\,, r=\frac{1}{l_u}\right\},
\end{align}
which is just the curve found in \cite{Song:2016gtd} that related to the holographic entanglement entropy. This is exactly the \textit{observation 1} we made from Rindler method.

Also it is easy to see that, $\mathcal{N}_\pm$ intersect with the asymptotic boundary on $u=\pm\frac{l_u}{2}$, thus enclose a strip $-\frac{l_u}{2}<u<\frac{l_u}{2}$ on the boundary, which is just the strip region $\mathcal{D}$ \eqref{mathcalD}. This confirms our \textit{observation 3} if $\mathcal{N}_\pm$ are normal to $\mathcal{E}$ (see section \ref{secmodularflows}).

According to the logic of Rindler method, the thermal entropy for $\widetilde{\text{AdS}_{3}}$ gives the entanglement entropy of $\mathcal{A}$. Since the thermal entropy is infinite, we need to regulate the interval $\mathcal{A}$ by a cutoff $\epsilon_{u}$ along the $u$ direction, such that 
\begin{align}\label{Areg}
\mathcal{A}_{reg}:~~\left\{(-\frac{l_u}{2}+\epsilon_u,-\frac{l_v}{2})\to(\frac{l_u}{2}-\epsilon_u,\frac{l_v}{2})\right\}\,.
\end{align}
As a consequence the extension of the horizon in Rindler $\widetilde{\text{AdS}_{3}}$ and the curve $\mathcal{E}$ in the original AdS$_3$ are also regulated.  We find the regulated $\mathcal{E}$ is given by \cite{Song:2016gtd}
\begin{align}\label{EA}
\mathcal{E}_{reg}:~\left\{(u,v,r)|~u=0,~r=\frac{1}{l_u},~v=\ell\left(l_v+\log\frac{l_u}{\epsilon_{u}}\right)(\eta-\frac{1}{2}),~~~~\eta\in[0,1]\right\}\,.
\end{align}
We see that $\mathcal{E}$ is cut off in the bulk at a finite radius $r=\frac{1}{l_u}$, rather than the asymptotic boundary. The holographic entanglement entropy is then given by
\begin{align}\label{Seerindler}
S_{\text{EE}}&=\frac{Length(\mathcal{E}_{reg})}{4G}=\frac{1}{4G}\left(l_v+\log\frac{l_u}{\epsilon_{u}}\right)\,.
\end{align}
Note that there is no need to introduce the cutoff $\epsilon_v$ along the $v$ direction since it can be taken to be zero without introducing extra divergence to the entanglement entropy.

Before going on, let us comment on the case of WAdS$_3$/WCFT. As was pointed out in \cite{Song:2016gtd}, the Rindler method applied on warped AdS$_3$ is indeed the same as the above story on AdS$_3$, because the warping factor appears in neither the Rindler transformations nor the thermodynamic quantities. For simplicity we only focus on the case of AdS$_3$/WCFT.

\section{Modular flows and modular planes}\label{secmodularflows}
The Rindler method can also help us find the explicit formula for the modular flow.  The generator of the normal Hamiltonian in Rindler space or Rindler bulk, which maps to the modular Hamiltonian in the original space, is the generator along the thermal circle, i.e. $k_t\equiv {\tilde \beta}^i \p_{\tilde{x}^i}$. Since $\partial_{\tilde{x}^i}$ can be written as a linear combination of the global generators, $k_t$ should have the same property. Using the holography dictionary, we can easily get the bulk dual of $k_t$, which we call $k_t^{bulk}$. In order to map it to the original space, we need to solve the following differential equations
\begin{align}\label{diffeq}
\partial_u&=(\partial_u \tilde{u})\partial_{\tilde{u}}+(\partial_u \tilde{v})\partial_{\tilde{v}}+(\partial_u \tilde{r})\partial_{\tilde{r}}\,,
\cr
\partial_v&=(\partial_v \tilde{u})\partial_{\tilde{u}}+(\partial_v \tilde{v})\partial_{\tilde{v}}+(\partial_v \tilde{r})\partial_{\tilde{r}}\,,
\cr
\partial_r&=(\partial_r \tilde{u})\partial_{\tilde{u}}+(\partial_r \tilde{v})\partial_{\tilde{v}}+(\partial_r \tilde{r})\partial_{\tilde{r}}\,.
\end{align}
So we can get $\partial_{\tilde{u}},\partial_{\tilde{v}},\partial_{\tilde{r}}$, and furthermore $k_t^{bulk}$, in terms of $\partial_u,\partial_v,\partial_r$.

We plug \eqref{tutvtr} into \eqref{diffeq}. Solving the equations we get the bulk and boundary modular flow
\begin{align}\label{bulkflow}
k_t^{bulk}&=-\tilde{\beta}_{\tilde{u}}\partial_{\tilde{u}}+\tilde{\beta}_{\tilde{v}}\partial_{\tilde{v}}=\pi\left(\partial_{\tilde{v}}-\partial_{\tilde{u}}\right)
\cr
&=\frac{\pi  \left(r^{-2}+4 u^2-l_u^2\right)}{2 l_u}\partial_u+\left(\pi -\frac{\pi }{l_u r}\right)\partial_v-\frac{4 \pi  r u}{l_u}\partial_r\,,
\\\label{boundaryflow}
k_t&=\frac{\pi  \left(4 u^2-l_u^2\right)}{2 l_u}\partial_u+\pi\partial_v\,.
\end{align}
It is easy to check that the curve $\mathcal{E}$ \eqref{gammawads} we find by Rindler method can also be determined by
\begin{align}
k^{bulk}_{t}(\mathcal{E})=0\,.
\end{align}
This means $\mathcal{E}$ is the fixed points of $k^{bulk}_{t}$ (or the bulk replica symmetry) and confirms our \textit{observation 2}. Obviously the endpoints of $\mathcal{A}$ are neither the fixed points of $k_t$ nor $k_t^{bulk}$. From the modular flow point of view, this means $\mathcal{E}$ should not be anchored on $\partial\mathcal{A}$, thus break the homology constraint.

\begin{figure}[h] 
   \centering
        \includegraphics[width=0.4\textwidth]{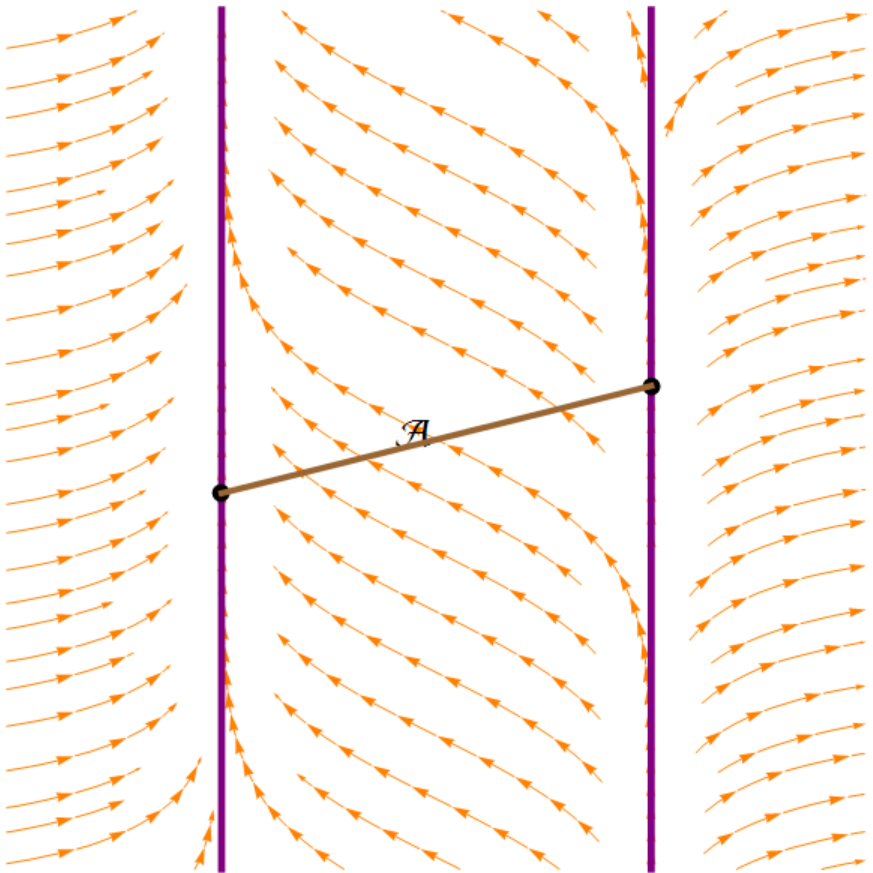}  
 \caption{ The orange lines with arrows depict the trajectory of the modular flow in WCFT. Note that the modular flow can never pass through $\partial\mathcal{D}$, which is depicted by the two purple lines.
\label{modularflowwcft} }
\end{figure}

We can get the explicit picture of the flow from \eqref{bulkflow} and \eqref{boundaryflow}. Solving the equation $(\frac{du(s)}{ds},\frac{dv(s)}{ds})=k_t$ we get the lines along modular flow on the boundary (see Fig.\ref{modularflowwcft}),
\begin{align}\label{vuboundary}
 \mathcal{L}_{v_0}:~~v=v_0-\text{arctanh}\frac{2 u}{l_u}\,,
 \end{align} 
where $v_0$ is a integration constant that characterizes different modular flow lines. It is easy to see that $\mathcal{L}_{v_0}$ is anti-symmetric with respect to its middle point $(0,v_0)$.

Similarly, by solving the equation $(\frac{du(s)}{d s},\frac{dv(s)}{d s},\frac{dr(s)}{d s})=k_t^{bulk}$ we get the functions of the bulk modular flow lines $\bar{\mathcal{L}}_{\bar v_0}^{r_0}$ that are described by the following two branches of solutions 
\begin{align}\label{branchA}
branch\,A&:~~\Big\{
\begin{array}{cc}
&u(r)=-\frac{1}{2 r}\sqrt{\frac{(r-r_0) \left(l_u^2 r r_0-1\right)}{r_0}}\,,
\\
&v(r)=\bar v_0+\frac{1}{2} \log \left(\frac{1+l_u r+\sqrt{(r-r_0) \left(l_u^2 r r_0-1\right)/r_0}}{1+l_u r-\sqrt{(r-r_0) \left(l_u^2 r r_0-1\right)/r_0}}\right) \,,\\ 
  \end{array} \quad -\frac{l_u}{2} < u \leq 0 \,,
\\\label{branchB}
branch\,B&:~~\Big\{
\begin{array}{cc}
&u(r)=\frac{1}{2 r}\sqrt{\frac{(r-r_0) \left(l_u^2 r r_0-1\right)}{r_0}}\,,
\\
&v(r)=\bar v_0-\frac{1}{2} \log \left(\frac{1+l_u r+\sqrt{(r-r_0) \left(l_u^2 r r_0-1\right)/r_0}}{1+l_u r-\sqrt{(r-r_0) \left(l_u^2 r r_0-1\right)/r_0}}\right)\,,\\ 
  \end{array} \quad ~~0\leq u < \frac{l_u}{2}\,.
\end{align}
The constants $r_0$ and $\bar v_0$ are the integration constants characterizing different bulk modular flow lines. The A branch part and B branch part smoothly connected at the point $(u,v,r)=(0,\bar v_0,r_0)$, which is the turning point of $\bar{\mathcal{L}}_{\bar v_0}^{r_0}$.

\begin{figure}[h] 
   \centering
        \includegraphics[width=0.42\textwidth]{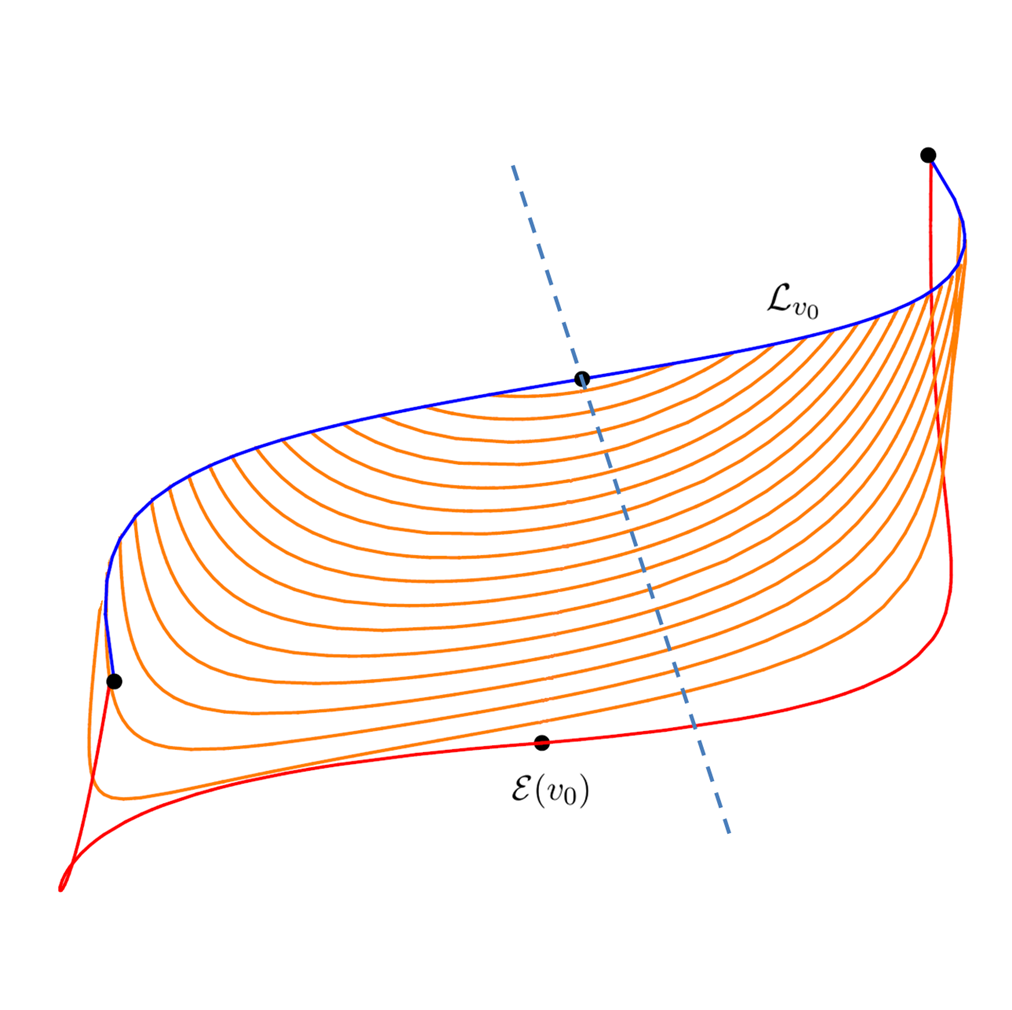}\quad
        \includegraphics[width=0.42\textwidth]{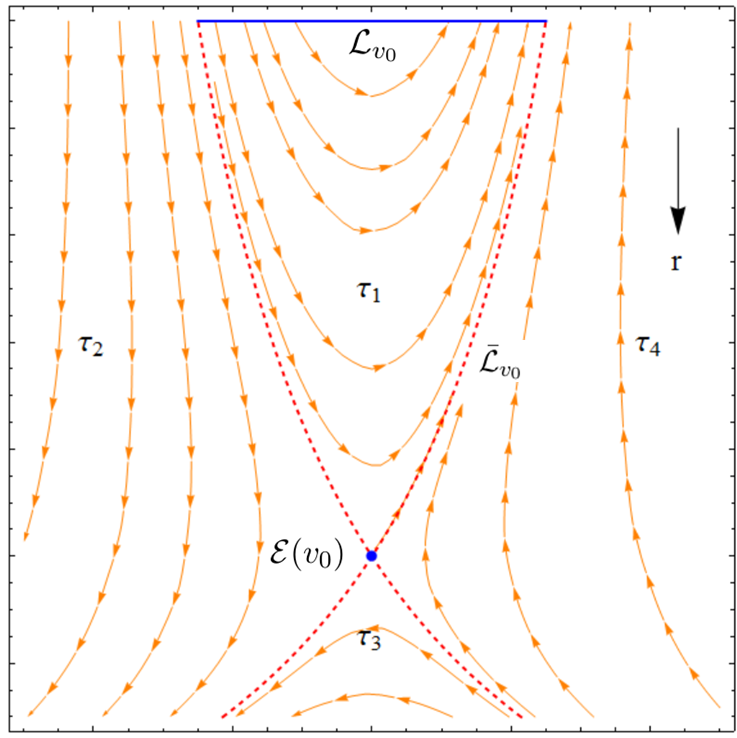}  
 \caption{The left figure gives an explicit diagram for a modular plane $\mathcal{P}(v_0)$. The blue line is $\mathcal{L}_{v_0}$ while the orange lines depict $\bar{\mathcal{L}}_{\bar{v}_0}^{r_0}$ with $\bar v_0=v_0\,,r> \frac{1}{l_u}$. The red line is $\bar{\mathcal{L}}_{v_0}$ with its turning point denoted as $\mathcal{E}(v_0)$. The other two black points are where $\mathcal{L}_{v_0}$ intersect with the red line at $v=\pm\infty$. The right figure is just the projection of the left figure to a flat plane. \label{modularp} }
\end{figure}

With the explicit picture of bulk and boundary modular flows, following \cite{Wen:2018whg} we then define the geometric quantity which we call the modular plane. When $r\to\infty$, all $\bar{\mathcal{L}}_{\bar{v}_0}^{r_0}$ will anchor on the two lines $u=\pm\frac{l_u}{2}$ (or $\partial\mathcal{D}$) at the boundary. However when we push the boundary into the bulk a little, the class of $\bar{\mathcal{L}}_{\bar{v}_0}^{r_0}$ with $\bar{v}_0=v_0$, will intersect with the boundary on $\mathcal{L}_{v_0}$. This class of bulk modular flow lines forms a codimension one surface in the bulk, which we call the modular plane $\mathcal{P}(v_0)$. In other words,
\begin{itemize}
\item \textit{the modular plane $\mathcal{P}(v_0)$ is the orbit of the boundary modular flow line $\mathcal{L}_{v_0}$ under the bulk modular flow.}
\end{itemize}
The modular planes are in one-to-one correspondence with the boundary modular flow lines.  See Fig.\ref{modularp} for an explicit diagram for a modular plane. Later we will show that the modular planes play a crucial role on how we regulate $\mathcal{E}$ in the bulk.

Another class of bulk modular flow lines are those with $r_0=\frac{1}{l_u}$. The turning points are just the fixed points of $k_t^{bulk}$ on $\mathcal{E}$. It is easy to check that these modular flow lines are null and normal to $\mathcal{E}$, thus form the two normal null hypersurfaces $\mathcal{N}_{\pm}$ \eqref{lightsheetwads} emanating from $\mathcal{E}$. We denote this class of bulk modular flow lines as $\bar{\mathcal{L}}_{\bar v_0}$
\begin{align}\label{nullgeodesics}
\bar{\mathcal{L}}_{\bar v_0}&:~~\Big\{
\begin{array}{cc}
&u=\frac{1}{2} \left(\frac{1}{r}-l_u\right),\qquad ~~v=\bar v_0+\frac{1}{2} \log (l_u r)\,,\qquad branch~~A\,,
\\
&u=-\frac{1}{2} \left(\frac{1}{r}-l_u\right),\qquad v=\bar v_0-\frac{1}{2} \log (l_u r)\,,\qquad branch~~B\,. 
  \end{array} 
\end{align}

\section{Generalized gravitational entropy for AdS$_3$ with CSS boundary conditions}\label{secgeneralizedge}
In this section we try to understand how the LM prescription \cite{Lewkowycz:2013nqa,Dong:2016hjy} works in AdS$_3$ with the CSS boundary conditions. In the rest of this section, we will use the terminologies in \cite{Dong:2016hjy}. For simplicity we will not repeat the Schwinger-Keldysh construction or time-folded path integral as in \cite{Dong:2016hjy}, but calculate Tr$\rho_{\mathcal{A}}$ by performing the path integral over the whole spacetime. The generalization to the time-folded path integral can be obtained following the lines in \cite{Dong:2016hjy}. We also try to extend the replica story of the boundary field theory into the bulk, and assume the replica symmetry is unbroken in the bulk. We will use \eqref{causalconsistence} to determine the bulk extremal surface $\mathcal{E}$, and use the modular planes to relate the bulk and boundary cutoffs.

We will give the replica story for WCFT in subsection \ref{boundarystory}, then we construct the dual bulk replica story in subsection \ref{bulkstory}. Then in subsection \ref{finestructure} we follow the prescription in \cite{Wen:2018whg} to study the fine structure of the bulk story using the modular planes. With the fine structure known, we will show the UV and IR cutoffs can be naturally related by the modular planes. At last, based on the fine structure analysis, we propose an intrinsic prescription of construct the geometric picture of entanglement entropy in subsection \ref{5.4}.

\subsection{The replica story on the boundary}\label{boundarystory}
The field theory dual of AdS$_3$ \eqref{tu0tv1} with CSS boundary conditions is a WCFT with the following thermal circle
\begin{align}\label{tcircleuv}
 (u,v)\sim(u,v-\pi i )\,.
 \end{align} 
The Rindler transformations decompose $\mathcal{B}$ into several regions with Rindler coordinates $(\tilde{u}_i,\tilde{v}_i)$. Similar to the strategy in \cite{Dong:2016hjy} we can simultaneously refer to all these spacetime regions in question, by allowing the Rindler coordinates to be complex and assigning discrete imaginary parts to each region. The subscript $i$ denote different spacetime regions. Moving along the modular flow and jump from one region to another, we hop from one imaginary part to another. If eventually we jump back to our starting point, the imaginary parts we have gone through form a circle, which can be physically understood as the thermal circle measured by the Rindler observer. Putting all the regions together, whose imaginary parts form the total thermal circle, we should recover a pure state. 

We cut the WCFT open along $\mathcal{A}$ then glue all the copies cyclically. We also need to know how the thermal circle changes under this cyclical gluing. The key to understand this is the assignment of the imaginary parts to all the spacetime regions. Since WCFT is not a relativistic field theory, its causal structure is quite abnormal compared with the relativistic ones, hence deserves discussion in detail.

For convenience, here we re-write the boundary Rindler transformation \eqref{boundarymap}
\begin{align}\label{boundarymap2}
\tanh\left(\tilde{u}\right)=\frac{2 u}{l_u}\,,
\qquad
\tanh\left(\tilde{v}\right)=\tanh v\,.
\end{align} 
We write the second equation in such a way that the thermal circles in both of the original and Rindler space are exhibited in the coordinate transformation, i.e. \eqref{tcircleuv} and
\begin{align}
(\tilde{u},\tilde{v})\sim(\tilde{u}+\pi i,\tilde{v}-\pi i)\,.
\end{align}
Then it is convenient to confine 
\begin{align}\label{confinement}
0\leq \text{Im}(\tilde{u})\leq i\pi\,,\qquad -i\pi\leq \text{Im}(\tilde{v})\leq 0\,.
\end{align}
Since the Rindler transformations divide the original spacetime $\mathcal{B}$ into three spacetime regions, we require the assignment of imaginary parts should consist with the Rindler transformations. In other words the Rindler transformations should map the Rindler space $(\tilde{u}_i,\tilde{v}_i)$ with different imaginary parts to different spacetime regions on $\mathcal{B}$. Furthermore we require the imaginary part of each region should be unique under the confinement \eqref{confinement} and different from the other regions. 

Note that, the assignment is not uniquely determined by the above requirements. For example, for the strip region we can chose either the assignment $\text{Im}[(\tilde{u},\tilde{v})]=(0,0)$ or $\text{Im}[(\tilde{u},\tilde{v})]=(0,-i\pi)$.  Both of the choices consist with the Rindler transformations. However all the choices satisfying our requirements can be related by a rotation or reversion of the thermal circle, thus do not change the physical story. Here we choose the following assignment for the three regions on $\mathcal{B}$
\begin{align}\label{imtau2}
\text{Im}[(\tilde{u},\tilde{v})]=\Big\{
\begin{array}{cc}
~(0,0)\,,\qquad &\frac{l_u}{2}<u<\frac{l_u}{2}\,,
\\
~(\frac{\pi i}{2},0)\,,\qquad &u<-\frac{l_u}{2}\,,
\\
(\frac{\pi i}{2},-\pi i)\,,\qquad &u>\frac{l_u}{2}\,. 
\end{array} 
\end{align}

According to uniqueness requirement for the imaginary parts, there is no place on $\mathcal{B}$ for the assignment $\text{Im}[(\tilde{u},\tilde{v})]=(0,-\pi i)$. This can be understood in the following way. The state on $\mathcal{B}$ is already mixed, which indicates the complete thermal circle involves a region outside $\mathcal{B}$ that purifies $\mathcal{B}$. We denote this region as $\mathcal{B}^{c}$ and consider $\mathcal{B}\cup\mathcal{B}^{c}$ as the spacetime that recovers the pure state. So the proper assignment should be
\begin{align}\label{imtau3}
\mathcal{B}^{c}:~~~~\text{Im}[(\tilde{u},\tilde{v})]=(0,-\pi i)\,.
\end{align}
We will see that the above assignment is quite natural in the gravity side story.

\begin{figure}[h] 
   \centering
        \includegraphics[width=0.4\textwidth]{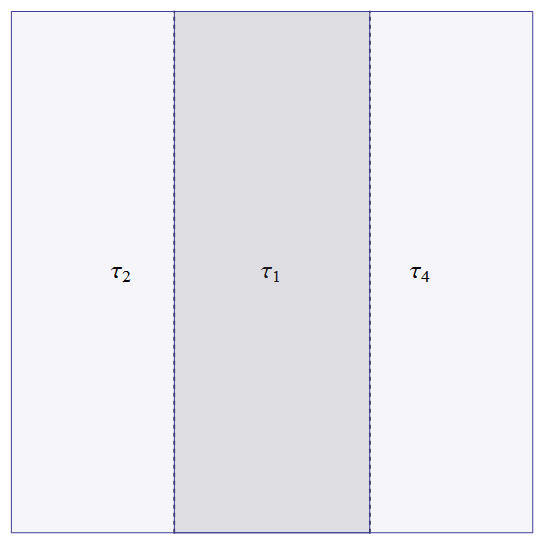} 
        \qquad
                \includegraphics[width=0.4\textwidth]{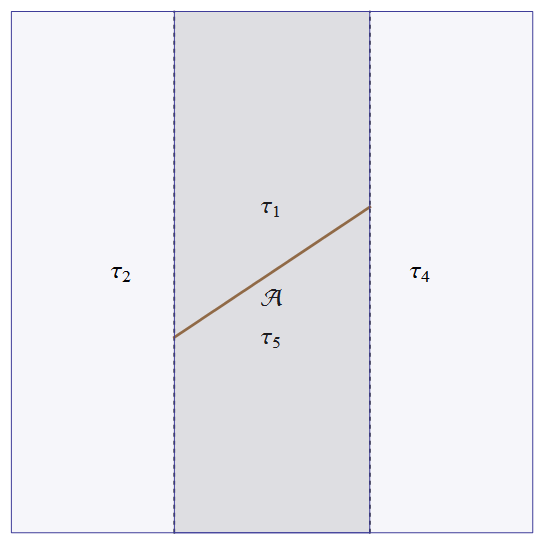} 
                \caption{The causal decomposition for the WCFT, and the assignment of the imaginary parts for each region.
\label{cd1} }
\end{figure}

It will be more convenient to introduce the ``Rindler time'' $\tau=\tilde{u}-\tilde{v}$ such that \begin{align}\label{imtau}
\tau_m=\tau+(m-1)\frac{\pi}{2}i\,.
\end{align}
The thermal circle becomes $\tau\sim\tau+2\pi i$ and the three regions on $\mathcal{B}$ can be denoted by $\tau_1,\tau_2$ and $\tau_4$ (see the left figure in Fig.\ref{cd1}), while the $\mathcal{B}^{c}$ can be denoted as $\tau_3$. For each time we cross the ``horizon'', we add a $\frac{\pi }{2}i$ to $\tau$. We cut $\mathcal{A}$ open into $\mathcal{A}_+$ and $\mathcal{A}_-$ then the strip region is further divided into $\tau_1,\tau_5$ (see the right figure in Fig.\ref{cd1}).

Then we construct the replica story on the field theory side. We consider n copies of $\mathcal{B}$ and glue them cyclically
\begin{align}\label{boundaryreplica}
\phi_I(\mathcal{A}_-)=\phi_{(I+1)}(\mathcal{A}_+)\,,\qquad \phi_n(\mathcal{A}_-)=\phi_{1}(\mathcal{A}_+)\,,\qquad I=1,\cdots, n-1\,.
\end{align} 
to calculate  Tr($\rho_\mathcal{A}^{n}$). After the gluing we get a $n$-sheet manifold $\mathcal{B}_{n}$ with replica symmetry, where we perform the path integral. For $n=1$, the $\tau_5$ region is glued back to the $\tau_1$ region at $\mathcal{A}$, thus the thermal circle on $\mathcal{B}$ is $\tau\sim\tau+2\pi i$. Similarly we find the thermal circle on $\mathcal{B}_{n}$ becomes $\tau\sim\tau+2\pi n i$. 

The fixed point of the thermal circle (or the replica symmetry) is where the thermal circle shrinks. More explicitly it should be the joint point of all the spacetime regions. However, unlike the case of CFT$_2$ (or other relativistic theories), there is no such point in WCFT. This is not surprising because $\mathcal{B}^{c}$ is outside $\mathcal{B}$ and $k_t$ \eqref{boundaryflow} is nonvanishing everywhere on $\mathcal{B}$. In other words we have the replica symmetry, but there is no fixed point for this symmetry on $\mathcal{B}_n$.

\subsection{The replica story in the bulk}\label{bulkstory}
In this subsection we try to construct the bulk extension of the boundary replica story. As in \cite{Lewkowycz:2013nqa,Dong:2016hjy}, we need to make the basic assumptions to extend the boundary replica story into the bulk. We assume the AdS$_3$/WCFT correspondence between the bulk and boundary theories and also we assume the replica symmetry can be extended into the bulk. According to \cite{Lewkowycz:2013nqa} the curve $\mathcal{E}$ that is fixed under the bulk replica symmetry is extremal. For a given boundary subregion $\mathcal{A}$ and its causal development $\mathcal{D}_{\mathcal{A}}$, we determine the corresponding $\mathcal{E}$ by requiring that $\mathcal{E}$ and its null normal hypersurfaces $\mathcal{N}_{\pm}$ should satisfy \eqref{causalconsistence}. The $\mathcal{N}_{\pm}$ decompose the bulk into four regions denoted by $\tau_m$. Here $\tau_m$ parametrizes the bulk modular flows in each region. 

Following the above strategy we search the geodesics in the bulk (for details see appendix \ref{geodesics}) and find that the geodesic satisfying \eqref{causalconsistence} exists and is just  the curve $\mathcal{E}$ \eqref{gammawads} we found by Rindler method. This is not surprising since the $\mathcal{N}_{\pm}$ \eqref{lightsheetwads} are formed by the normal null geodesics associated to $\mathcal{E}$ and intersect with the boundary at $\partial\mathcal{D}_{\mathcal{A}}:~u=\pm\frac{l_u}{2}$. The bulk subregion enclosed by $\mathcal{B}$ and $\mathcal{N}_{\pm}$ is the analogue of the entanglement wedge $\mathcal{W}_{\mathcal{A}}$. 

\begin{figure}[h] 
   \centering
        \includegraphics[width=0.7\textwidth]{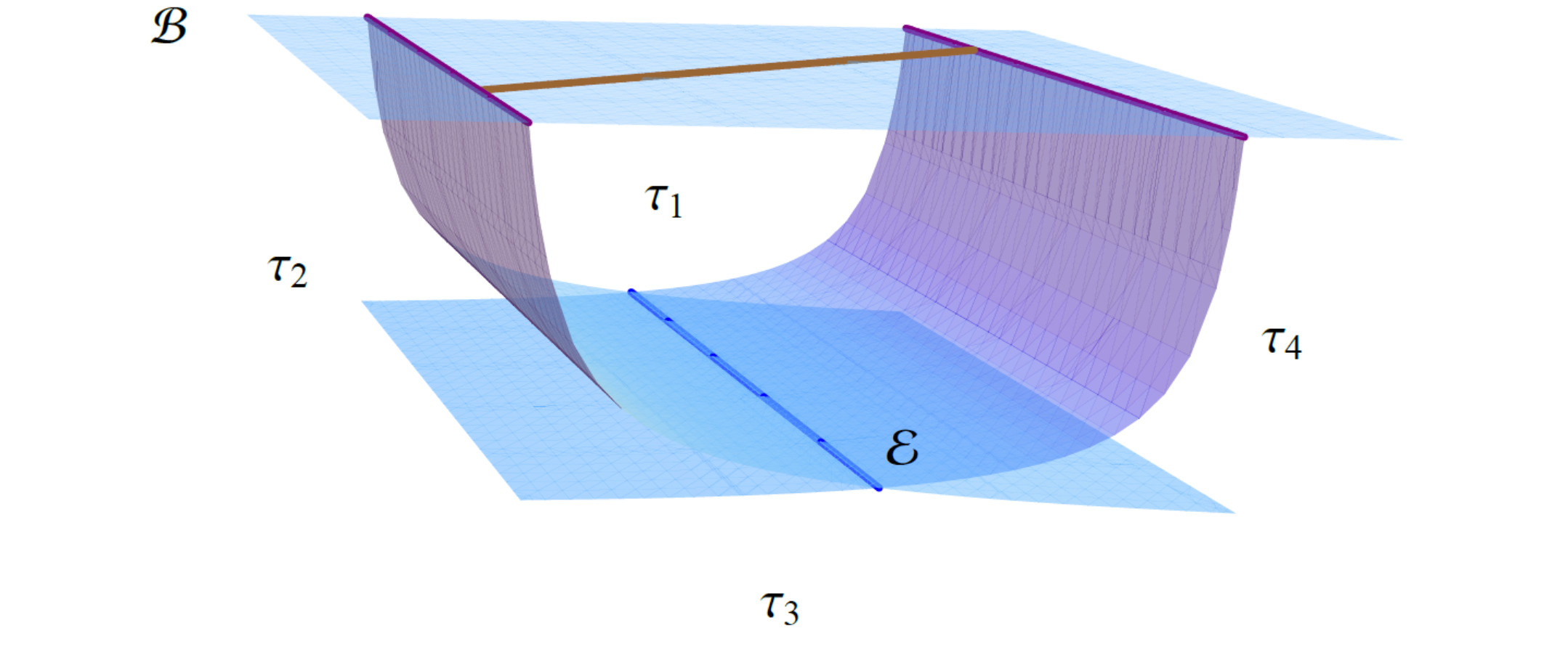}  
 \caption{ This figure shows the causal decomposition of the bulk $\mathcal{M}$ and the boundary $\mathcal{B}$. The two surfaces that intersect at $\mathcal{E}$ are $\mathcal{N}_{\pm}$ and the brown line is the boundary interval $\mathcal{A}$.
\label{bulk} }
\end{figure}

Now we try to construct the bulk replica story. We allow $\tau_m$ to be complex and refer to all the bulk regions in question by defining $\tau_m=\tau+\frac{(m-1)}{2}i\pi$. The assignment{\footnote{The assignment in the bulk can also be obtained from the bulk Rindler transformations, see also \cite{Sorkin:1974pya,Neiman:2013ap} for discussions related to the assignment.}} of the imaginary parts are explicitly shown in Fig.\ref{bulk}. Note that the $\tau_3$ region in the bulk does not overlap with the boundary $\mathcal{B}$, thus confirms our statement that there is no $\tau_3$ region on $\mathcal{B}$. We expect the boundary branched cover structure inherent in the replica construction to be inherited by the holographic map in the bulk. The bulk geometry should be a replicated geometry glued cyclically from $n$ copies of the bulk spacetime. 

Before we go ahead, we briefly review the bulk replica story in AdS/CFT \cite{Dong:2016hjy}. In this case the RT surface $\mathcal{E}$ is anchored on $\partial\mathcal{A}$. We denote the spacelike codimension one surface enclosed by $\mathcal{E}$ and $\mathcal{A}$ as $\mathcal{R}_\mathcal{A}$, which satisfies $\partial\mathcal{R}_{\mathcal{A}}=\mathcal{A}\cup\mathcal{E}$. Firstly, for each copy of bulk $\mathcal{M}^I$, we cut them open along $\mathcal{R}^{I}_\mathcal{A}$ into $\mathcal{R}^{I}_\mathcal{A}{}_+$ and $\mathcal{R}^{I}_\mathcal{A}{}_-$. Then we get the replicated geometry by gluing the open cuts cyclically
\begin{align}
\mathcal{R}^{I}_\mathcal{A}{}_-=\mathcal{R}^{(I+1)}_\mathcal{A}{}_+\,,\qquad \mathcal{R}^{n}_\mathcal{A}{}_-=\mathcal{R}^{1}_\mathcal{A}{}_+\,.
\end{align}

In the case of AdS/WCFT, in order to conduct the replica trick we also need to cut the bulk open along some codimension one surface, which we also denote as $\mathcal{R}_{\mathcal{A}}$.  Since $\mathcal{E}$ is supposed to be fixed under the replica symmetry, we require $\mathcal{E}\subset \partial\mathcal{R}_{\mathcal{A}}$. On the boundary, we require  $\mathcal{A}\subset \partial\mathcal{R}_{\mathcal{A}}$ to reproduce the boundary replica story. However, in this case $\mathcal{E}$ is not anchored on $\partial\mathcal{A}$, so $\partial\mathcal{R}_\mathcal{A}$ contains other parts. In addition, $\mathcal{R}_{\mathcal{A}}$ have two more boundaries $\gamma_\pm$ that connect the two endpoints $\partial\mathcal{A}_{\pm}$ of the boundary interval and the endpoints of $\mathcal{E}$ at $v=\pm\infty$. Later we will argue that $\gamma_{\pm}$ should be the null geodesics on $\mathcal{N}_{\pm}$. In summary we have
\begin{align}\label{boundaryRA}
\partial\mathcal{R}_{\mathcal{A}}=\mathcal{A}\cup\mathcal{E}\cup\gamma_+\cup\gamma_-\,.
\end{align}
With $\partial\mathcal{R}_{\mathcal{A}}$ settled down, the surface $\mathcal{R}_{\mathcal{A}}$ has the freedom to vibrate as long as we keep $\mathcal{R}_{\mathcal{A}}$ spacelike everywhere except $\gamma_{\pm}$. Then we cut $\mathcal{R}_{\mathcal{A}}^{I}$ open to $\mathcal{R}_{\mathcal{A}+}^{I}$ and $\mathcal{R}_{\mathcal{A}-}^{I}$ in each copy of the bulk and  then glue them cyclicly to the replicated geometry $\mathcal{M}_{n}$.

\begin{figure}[h] 
   \centering
        \includegraphics[width=0.5\textwidth]{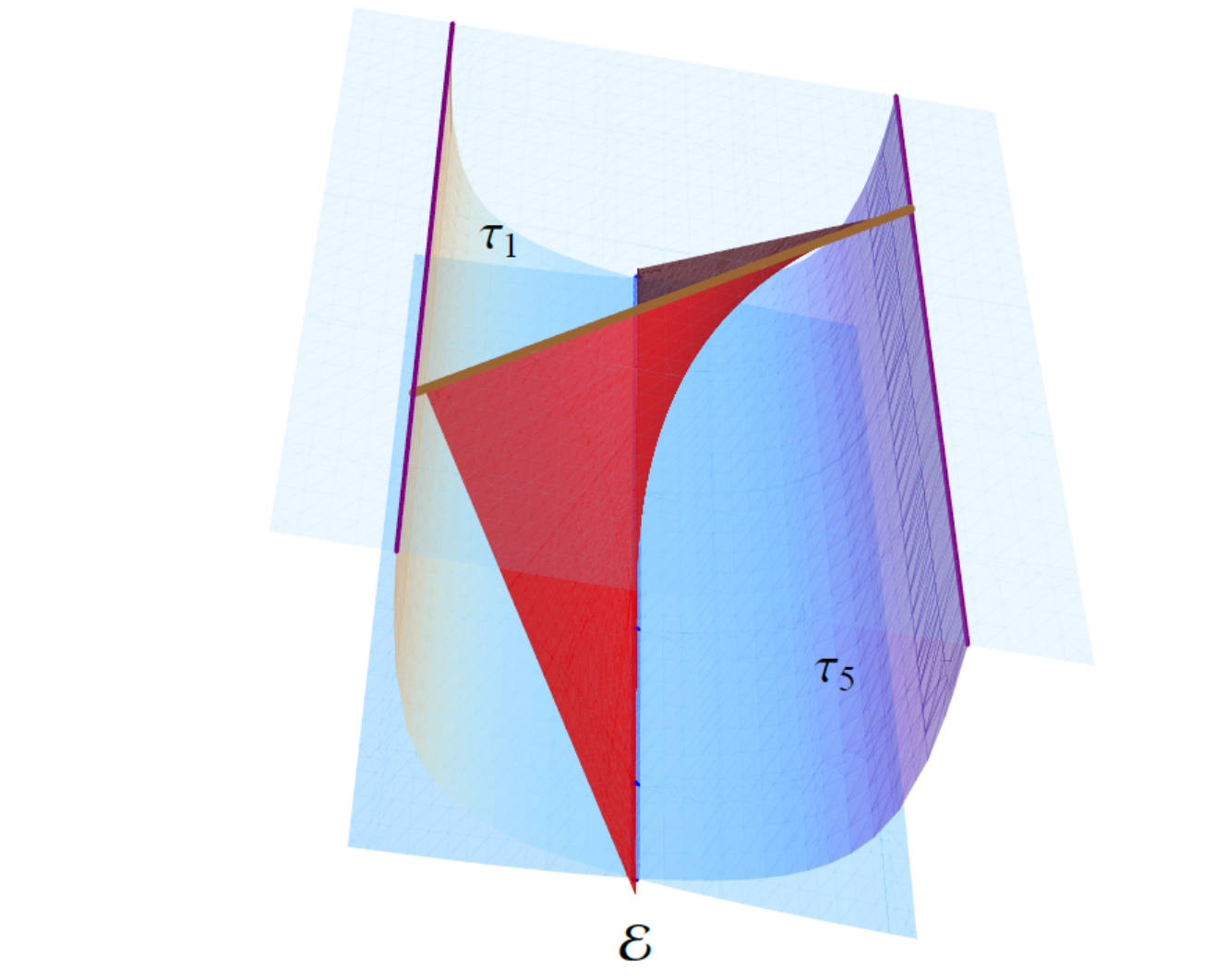}  
 \caption{ The red surface is the $\mathcal{R}_{\mathcal{A}}$ where we cut the bulk open.
\label{bulk2} }
\end{figure}

Note that when we cut $\mathcal{R}_{\mathcal{A}}$ open we divide the $\tau_1$ region into two regions denoted by $\tau_1$ and $\tau_5$ (see Fig. \ref{bulk2}). Starting from some point in the $\tau_1$ region, we move along a bulk modular flow line and cross the horizons $\mathcal{N}_{\pm}$ for four times to arrive at the $\tau_5$ region. Then we pass through $\mathcal{R}_{A}$ and enter the next copy of bulk spacetime. The cyclic gluing of $n$ copies of the bulk makes us pass through the horizons for 4$n$ times to get back to the starting point. This induces the thermal circle $\tau\sim\tau+2\pi n i$ in $\mathcal{M}_{n}$, which shrinks at $\mathcal{E}$. As expected the bulk replica story reproduces the boundary replica story.

\subsection{Relating the UV and IR cutoffs with the modular planes (null geodesics)}\label{finestructure}

Then we focus on our second task: how to regulate $\mathcal{E}$ when we regulate the boundary interval $\mathcal{A}$. 
In the case of AdS/CFT, the UV/IR relation \cite{Susskind:1998dq} is used to regulate the holographic entanglement entropy in the RT formula. This prescription for regulation is also confirmed in \cite{Wen:2018whg} by using the modular planes as a slicing of the entanglement wedge. The slicing gives a fine correspondence between the points on $\mathcal{A}$ and the points on the RT surface $\mathcal{E}$. \cite{Wen:2018whg} shows that the points where we cut off $\mathcal{A}$ and $\mathcal{E}$ should be related by this fine correspondence.

In the case of AdS$_3$/WCFT, the curve $\mathcal{E}$ \eqref{gammawads} can not be regulated following the RT formula. However, with the bulk and boundary modular flows clear, we can also define the modular planes as in \cite{Wen:2018whg} and perform the fine structure analysis to determine the cutoff point in the bulk. As in the case of AdS$_3$/CFT$_2$ \cite{Wen:2018whg}, we will show that the modular planes can regulate $\mathcal{E}$ correctly when we regulate $\mathcal{A}$.

The modular plane $\mathcal{P}(v_0)$ is defined as the orbit of a boundary modular flow line $\mathcal{L}_{v_0}$ under the bulk modular flow, which is a codimension one surface in the bulk. Its construction in this case is discussed in details in section \ref{secmodularflows}. The modular planes are in one-to-one correspondence with the boundary modular flow lines $\mathcal{L}_{v_0}$. By definition we have
\begin{align}
\mathcal{P}(v_0)\cap\mathcal{B}=\mathcal{L}_{v_0}\,.
\end{align}
As the boundary can be viewed as a slicing of modular flow lines $\mathcal{L}_{v_0}$, the entanglement wedge $\mathcal{W}_{\mathcal{A}}$ can also be viewed as a slicing of the modular planes $\mathcal{P}(v_0)$. The normal null geodesics on $\mathcal{N}_\pm$ are also bulk modular flow lines, which indicates the modular planes will intersect with $\mathcal{N}_{\pm}$ on these normal null geodesics, i.e. 
\begin{align}
\mathcal{P}(v_0)\cap\mathcal{N}_{\pm}=\bar{\mathcal{L}}_{v_0}\,.
\end{align}
It is easy to see that $\bar{\mathcal{L}}_{v_0}$ (or $\mathcal{P}(v_0)$) intersect with $\mathcal{E}$ \eqref{gammawads} at its turning point
\begin{align}
\mathcal{P}(v_0)\cap \mathcal{E}= \mathcal{E}(v_0):~(u,v,r)=(0,v_0,\frac{1}{l_u})\,.
\end{align} 
Also each modular plane will intersect with $\mathcal{R}_{\mathcal{A}}$ on a line
\begin{align}
\mathcal{R}_{\mathcal{A}}(v_0)=\mathcal{P}(v_0)\cap\mathcal{R}_{\mathcal{A}}\,.
\end{align} 
See Fig.\ref{modularp} for a typical modular plane.

\begin{figure}[h] 
   \centering
        \includegraphics[width=0.8\textwidth]{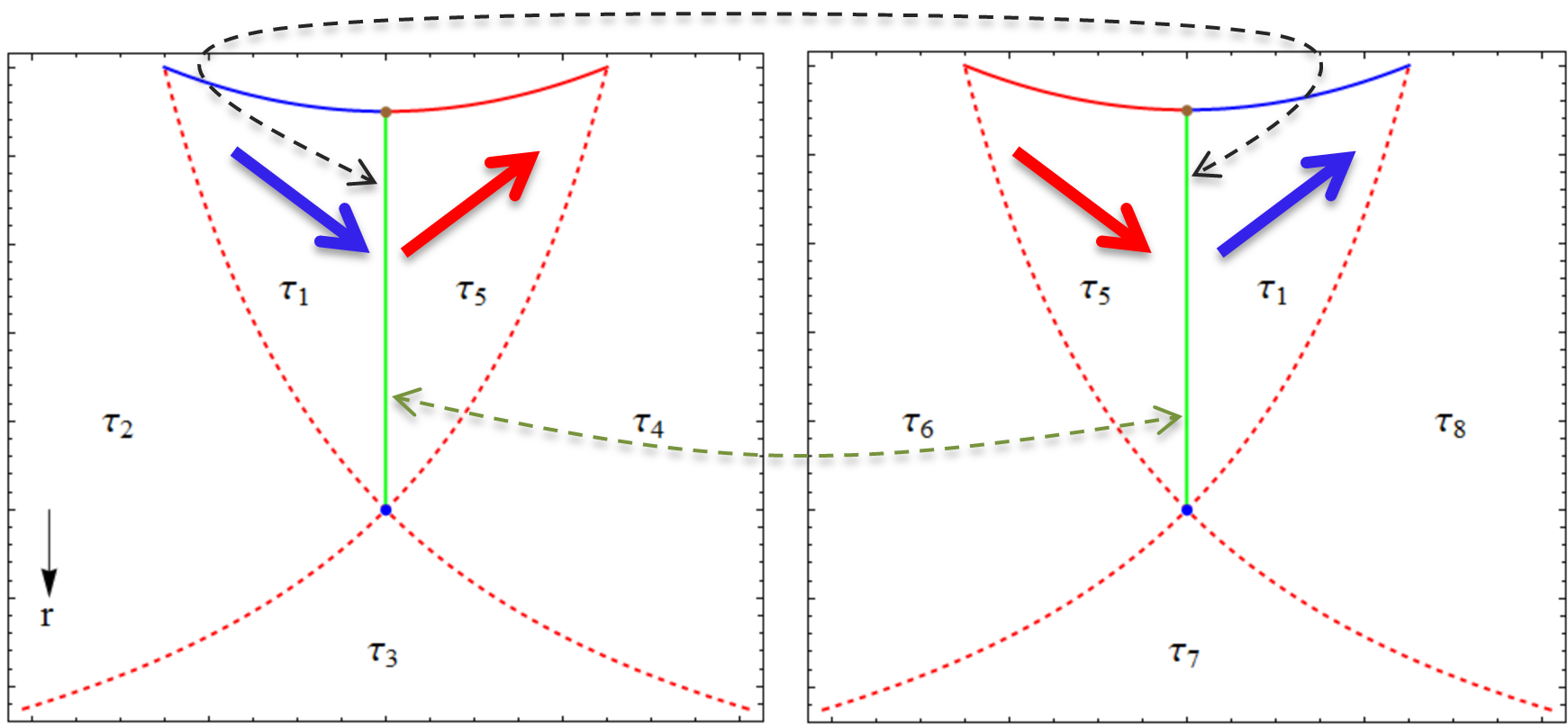}  
 \caption{ This figure is taken from \cite{Wen:2018whg} and shows the replica story for a cyclically glued modular plane $\mathcal{P}_{n}(v_0)$ with $n=2$. In the first copy the boundary modular flow line $\mathcal{L}_{v_0}$ parametrized by $\tau_1$ (blue line) passes through $\mathcal{A}(v_0')$ and get into the next copy. The bulk flow (blue arrow) should also go through $\mathcal{R}_{\mathcal{A}}(v_0)$ (the green line) to the next copy then go back to $\mathcal{L}_{v_0}$. A similar story happens to the flow lines parametrized by $\tau_5$ (the red line and arrow). The dashed arrows shows the cyclic gluing on $\mathcal{R}_{\mathcal{A}}(v_0)^{1,2}$. 
\label{bulkreplica} }
\end{figure}

Translation along a bulk modular flow line is a translation of the real part of $\tau_{m}$ while keeping the imaginary part fixed. When we apply the replica trick, the bulk and boundary are cyclically glued, the orbit of the modular flow changes, as well as the distribution of the imaginary part Im[$\tau$]. Let us focus on the cyclic gluing of a single point $\mathcal{A}(v'_0)$ and see how it changes the modular flow picture both in the bulk and boundary. Here $v'_0$ denote the $v$ coordinate of the point $\mathcal{A}(v'_0)$. We denote the boundary modular flow line that passes this point as $\mathcal{L}_{v_0}$ given by \eqref{vuboundary}. On the boundary, $\mathcal{L}_{v_0}$ will enter the next copy of $\mathcal{B}$ when it passes through $\mathcal{A}(v'_0)$.  By definition all the bulk modular flow lines that emanating from $\mathcal{L}_{v_0}$ will return back to $\mathcal{L}_{v_0}$. As $\mathcal{L}_{v_0}$ enters the next copy of $\mathcal{B}$, the bulk modular flow emanating from $\mathcal{L}_{v_0}$ also need to enter the next copy of bulk to get back to the same $\mathcal{L}_{v_0}$. Then the natural bulk extension of the cyclic gluing of the point $\mathcal{A}(v'_{0})$ is the cyclic gluing of the modular plane $\mathcal{P}(v_0)$ on $\mathcal{R}_{\mathcal{A}}(v_0)$, i.e.
\begin{align}\label{replicaplane}
\mathcal{R}_{\mathcal{A-}}^{I}(v_0)&=\mathcal{R}_{\mathcal{A+}}^{I+1}(v_0)\,,\qquad \mathcal{R}_{\mathcal{A-}}^{n}(v_0)=\mathcal{R}_{\mathcal{A+}}^{1}(v_0)\,.
\end{align}
We denote the cyclically glued modular plane as $\mathcal{P}_{n}(v_0)$. Following the modular flows we can keep track of the imaginary part of $\tau$ everywhere on $\mathcal{P}_{n}(v_0)$ and find the thermal circle on $\mathcal{P}_{n}(v_0)$ becomes $\tau\sim\tau+2\pi n i$. Accordingly the induced metric near the fixed point $\mathcal{E}(v_0)$ on $\mathcal{P}_{n}(v_0)$ becomes 
\begin{align}
ds^2=n^2 d\rho^2-\rho^2 d\tau^2+\cdots,
\end{align}
where $\rho$ denotes the distance from $\mathcal{E}(v_0)$, and the dots means the higher order terms. See Fig.\ref{bulkreplica} for the replica story on the modular plane with $n=2$.  The whole bulk replica story can be considered as a slicing of the replica stories on all the modular planes.

\begin{figure}[h] 
   \centering
        \includegraphics[width=0.35\textwidth]{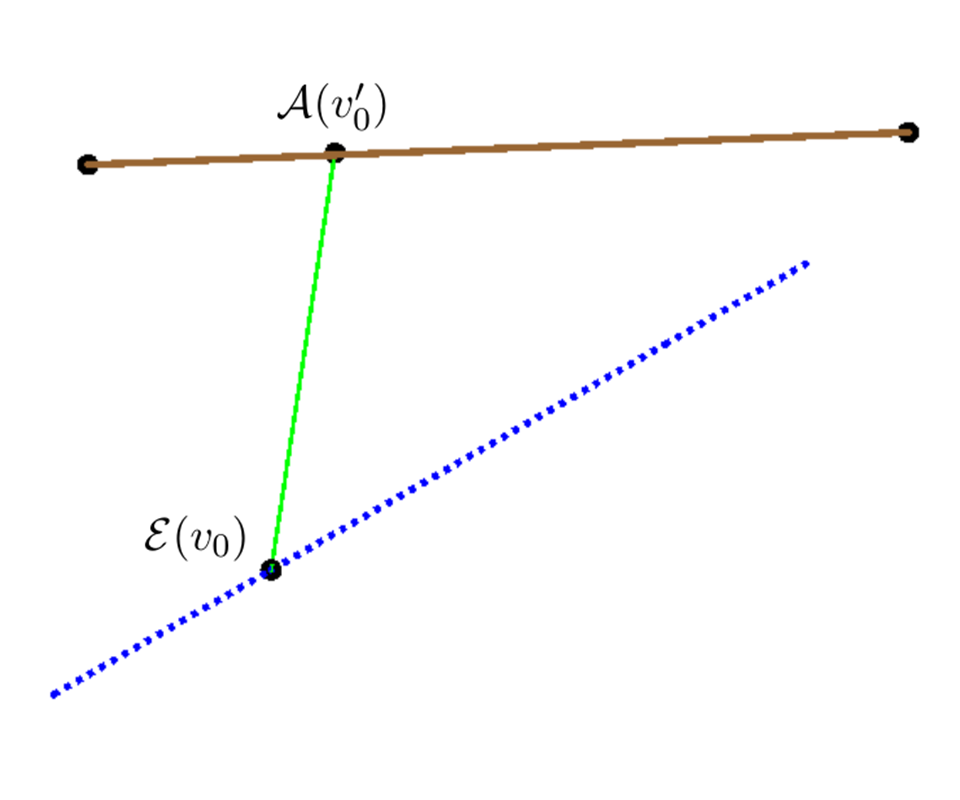} \qquad 
        \includegraphics[width=0.45\textwidth]{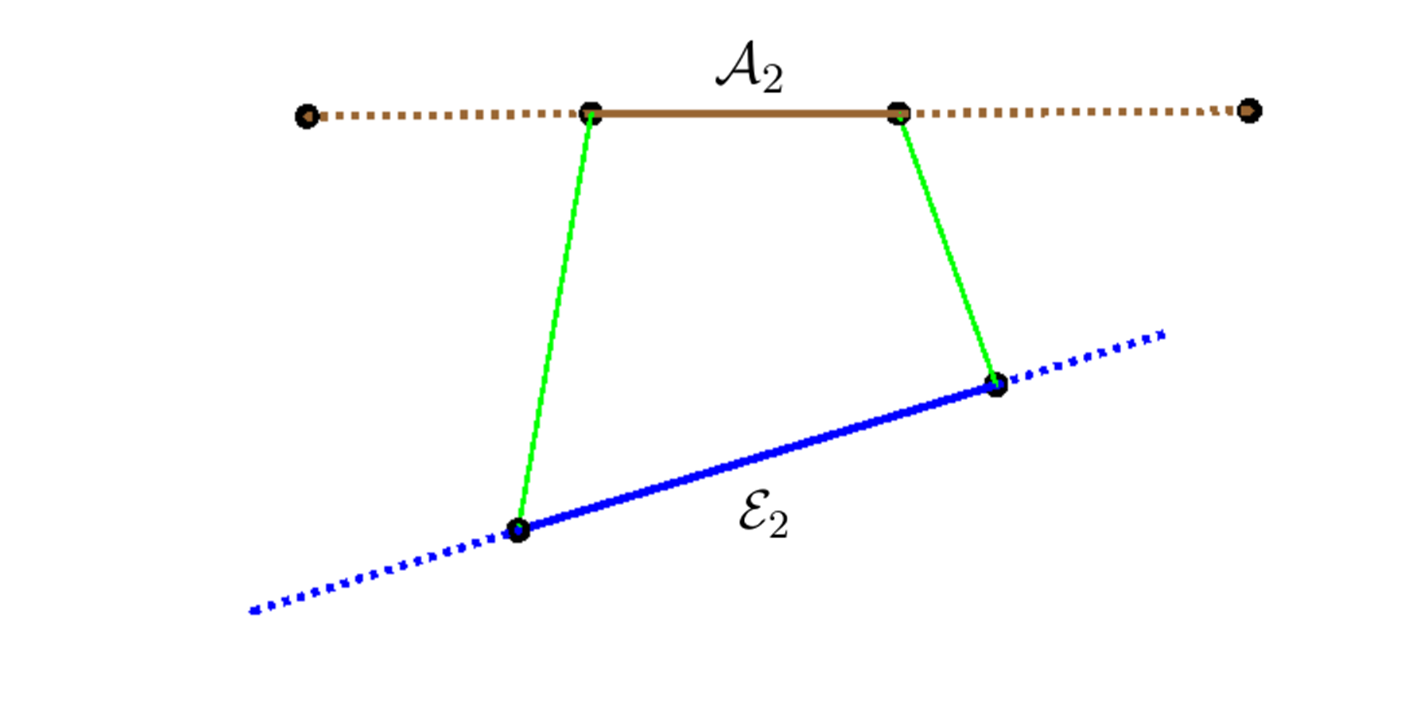}   
 \caption{The right figure shows the correspondence between the point $\mathcal{A}(v'_0)$ on $\mathcal{A}$ and its partner $\mathcal{E}(v_0)$ on $\mathcal{E}$ with $v_0$ and $v'_0$ satisfying \eqref{v0vp}.  Also they are the points where the modular plane $\mathcal{P}(v_0)$ intersect with $\mathcal{A}$ and $\mathcal{E}$. The green line is $\mathcal{R}_{\mathcal{A}}(v_0)$. The left figure shows, in the same sense, an arbitrary sub-interval $\mathcal{A}_2$ corresponds to a subinterval $\mathcal{E}_2$ on $\mathcal{E}$.
\label{pp1} }
\end{figure}

In summary, the cyclic gluing of a point $\mathcal{A}(v'_0)$ on the boundary interval induces a replica story on $\mathcal{P}_{n}(v_0)$. Following the calculations in \cite{Lewkowycz:2013nqa,Dong:2016hjy}, this turns on nonzero contribution to the entanglement entropy on the fixed point $\mathcal{E}(v_0)$ where the modular plane intersect with $\mathcal{E}$. In other words, this gives an one-to-one correspondence between the points $\mathcal{A}(v'_0)$ on $\mathcal{A}$ and the points $\mathcal{E}(v_0)$ on $\mathcal{E}$ (see the left figure of Fig.\ref{pp1}). More explicitly, if we consider $\mathcal{A}$ to be a straight line
\begin{align}\label{straightA}
 \mathcal{A}:~~ u=\frac{l_u}{l_v}v\,,\qquad-\frac{l_v}{2}\leq v \leq \frac{l_v}{2}\,,
 \end{align} 
the two points that correspond to each other are related by
\begin{align}\label{v0vp}
v'_0+\text{arctanh}\frac{2 v'_0}{l_v}=v_0\,.
\end{align}

When $\mathcal{A}(v'_0)$ approaches $\partial_{\pm}\mathcal{A}$, i.e. $v'_0=\pm\frac{l_v}{2}$, the partner points become $\mathcal{E}(v_0^{\pm})|_{v_0^{\pm}=\pm\infty}$. Note that the endpoints $\partial_{\pm} A$ also lie on the null hypersurfaces $\mathcal{N}_{\pm}$, they can be connected to their partners $\mathcal{E}(v_0^{\pm})$ by the two null geodesics $\gamma_{\pm}=\bar{\mathcal{L}}_{v_0^{\pm}}$. This indicates that all the lines that connect these two pair of points will contain timelike part except the two null lines $\gamma_{\pm}$. Since we do not expect time-like parts on the surface $\mathcal{R}_{\mathcal{A}}$, the only choice for $\mathcal{R}_{\mathcal{A}}(v_0^{\pm})$ are $\gamma_{\pm}$. This is how we determine $\partial\mathcal{R}_{\mathcal{A}}$ to be \eqref{boundaryRA}.

In the same sense, an arbitrary sub-interval $\mathcal{A}_2$ of $\mathcal{A}$ correspond to a sub-interval $\mathcal{E}_2$ on $\mathcal{E}$ (see the right figure of Fig.\ref{pp1}). We denote the $v$ coordinate of the two endpoints of $\mathcal{A}_2$ as $v'_1$ and $v'_2$ and  denote  the $v$ coordinates of the two endpoints of $\mathcal{E}_2$ as $v_1$ and $v_2$. Similarly they should be related by \eqref{v0vp}. According to our prescription, the contribution from $\mathcal{E}_2$ to the total entanglement entropy $S_{\mathcal{A}}$ is turned on by the cyclic gluing of $\mathcal{A}_2$ on the boundary. Thus it is natural to propose that the length of $\mathcal{E}_2$ captures the contribution from the sub-interval $\mathcal{A}_2$ to $S_{\mathcal{A}}$. We denote this contribution as $s_{\mathcal{A}}(\mathcal{A}_2)$ (see \eqref{SA22})
\begin{align}
s_{\mathcal{A}}(\mathcal{A}_2)=\frac{Length(\mathcal{E}_2)}{4G}\,.
\end{align}

\begin{figure}[h] 
   \centering
        \includegraphics[width=0.6\textwidth]{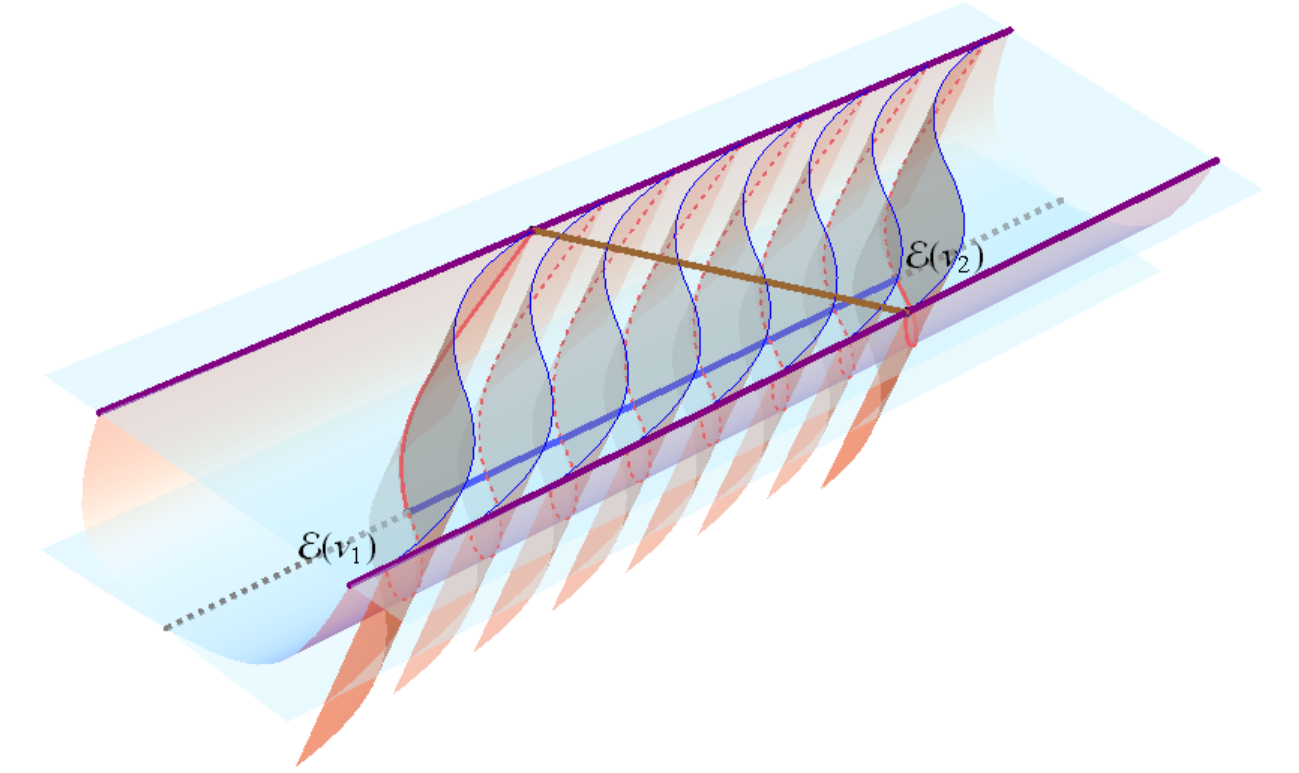}  
 \caption{ The entanglement wedge is shown as a slicing of the modular planes. We denote the thick blue line in the bulk as $\mathcal{E}_{reg}$. We only depicted the modular planes that go through $\mathcal{A}_{reg}$. The modular planes $\mathcal{P}(v_1)$ and $\mathcal{P}(v_2)$ intersect with $\mathcal{A}$ and $\mathcal{E}$ on the points where they are cut off.
\label{bulkrs} }
\end{figure}

Let us go a step further and consider $\mathcal{A}_2=\mathcal{A}_{reg}$. It is natural to interpret the regulated entanglement entropy as the contribution from $\mathcal{A}_{reg}$. More explicitly we set{\footnote{The additional terms proportional to $\epsilon_{u}$ in \eqref{vprime} appear because we regulate the $u$ direction with $\epsilon_u$ and keep the endpoints of $\mathcal{A}_{reg}$ on the straight line \eqref{straightA}.} }
\begin{align}\label{vprime}
 v'_1=-\frac{l_v}{2}+\epsilon_u\frac{l_v}{l_u}\,,\qquad v'_2=\frac{l_v}{2}-\epsilon_u\frac{l_v}{l_u}\,.
 \end{align} 
Then the end two points $\mathcal{E}(v_1)$ and $\mathcal{E}(v_2)$ of $\mathcal{E}_2=\mathcal{E}_{reg}$ should be the points where the geodesic $\mathcal{E}$ is cut off. Applying \eqref{v0vp}, we find
\begin{align}
v_1=-\frac{1}{2}\left(l_v+\log\frac{l_u}{\epsilon_u}\right)+\mathcal{O}(\epsilon_u)\,,\qquad v_2=\frac{1}{2}\left(l_v+\log\frac{l_u}{\epsilon_u}\right)+\mathcal{O}(\epsilon_u) \,,
\end{align}
which are exactly the cutoff points we found by Rindler method. For an overall picture of our prescription, see Fig.\ref{bulkrs}. The regulated entanglement entropy is then given by
\begin{align}\label{SeeLM}
S_{\mathcal{A}}=\frac{Length(\mathcal{E}_{reg})}{4G}=\frac{1}{4G}\left(l_v+\log\frac{l_u}{\epsilon_{u}}\right)\,.
\end{align}
As expected the result coincide with the result \eqref{Seerindler} we get by the Rindler method.

As was pointed out by \cite{Wen:2018whg}, the fine correspondence between the points on $\mathcal{A}$ and the points on $\mathcal{E}$ defines an entanglement contour function that describes the distribution of the entanglement on $\mathcal{A}$. People who are interested in the entanglement contour should consult appendix \ref{appendixA}, where we calculate the contour function based on this fine correspondence \eqref{v0vp}, and furthermore we test the proposal \cite{Wen:2018whg} for entanglement contour function for general theories.

\subsection{The intrinsic construction of the geometric picture}\label{5.4}
The construction above is concrete but relies heavily on the explicit picture of the bulk and boundary modular flows, which is complicated to calculate and only exists for special cases. Then we come to the important question: is there a prescription to construct the geometric picture intrinsically without the construction of Rindler transformations and the information of the modular flows? Inspired by the above construction we try to propose such a prescription for the case we study. 

The prescription also involves a cutoff at large radius $r_I$, which is related to the cutoff $\epsilon_u$ in the WCFT by
\begin{align}\label{UVIR}
r_I=\frac{1}{\epsilon_u}\,.
\end{align}
In this case the radius cutoff is imposed on the two null geodesics $\gamma_{\pm}$ rather than the spacelike geodesic $\mathcal{E}$. However the way we impose this cutoff is a little tricky. The $\gamma_\pm$ emanating from the boundary endpoints $\partial_{\pm} \mathcal{A}$ at the real boundary will intersect with $\mathcal{E}$ at $v_0^{\pm}=\pm\infty$. So cutting off $\gamma_\pm$ at $r_I$ does not regulate the entanglement entropy. 

\begin{figure}[h] 
   \centering
        \includegraphics[width=0.7\textwidth]{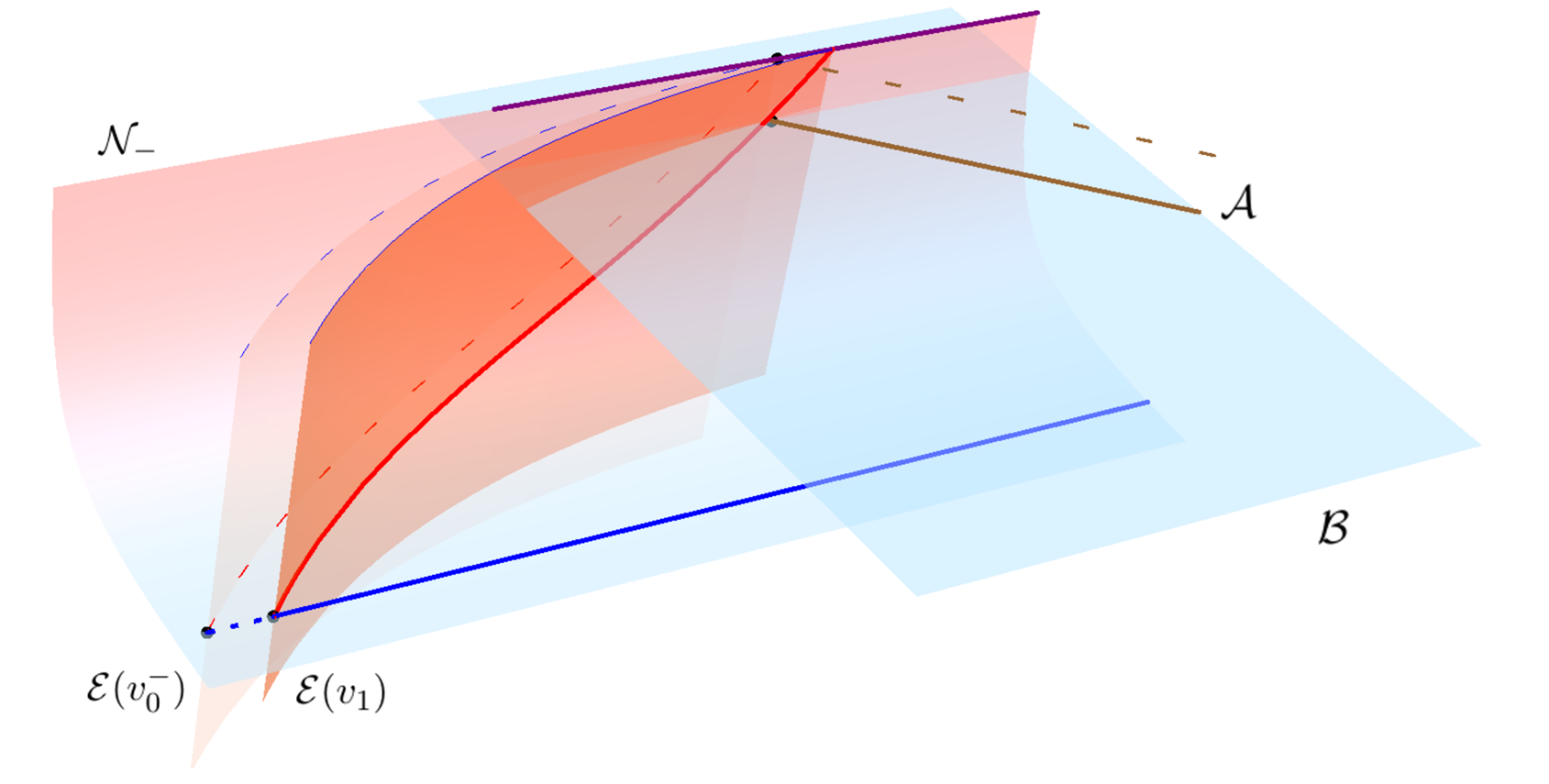}  
 \caption{We assume $\mathcal{E}(v_0^{-})$, with $v_0^-=-\infty$, is the point that connected to $\partial_-\mathcal{A}$ on the real boundary through the null geodesics $\gamma_-$. When we push $\partial_-\mathcal{A}$ to the cutoff boundary along $\mathcal{N}_-$, the null geodesic $\gamma_-$ that goes through it will change accordingly as well as the intersection point with $\mathcal{E}$. In such a way the curve $\mathcal{E}$ is regulated  through the null geodesics.
 \label{intrinsic2} }
\end{figure}

The right way to do the regulation is in the following. We first need to push the WCFT to the cutoff boundary at $r=r_I$. During we push the boundary, we should keep $\partial_\pm \mathcal{A}$ on $\mathcal{N}_{\pm}$ thus adapt to the bulk causal decomposition. Furthermore we should keep the $v$ coordinate of $\partial_\pm\mathcal{A}$ fixed since there is no cutoff along the $v$ direction. Then the two null geodesics that emanating from the $\partial_\pm\mathcal{A}$ at $r=r_I$ will intersect with $\mathcal{E}$ on the right cutoff points (see Fig. \ref{intrinsic2}).

\begin{figure}[h] 
   \centering
        \includegraphics[width=0.7\textwidth]{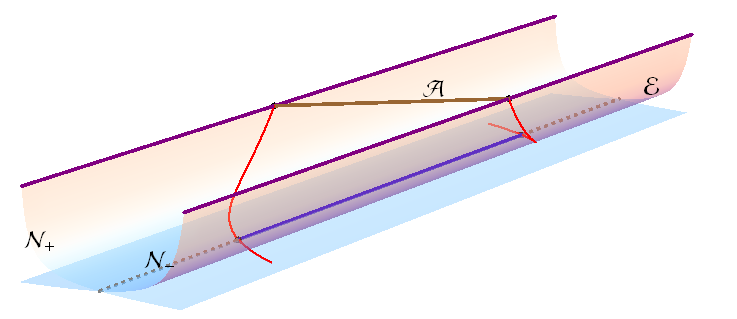}  
 \caption{In this figure the interval $\mathcal{A}$ is on the cutoff boundary $r=r_I$ and its endpoints $\partial_\pm \mathcal{A}$ are on $\mathcal{N}_{\pm}$. The two red lines are the two null geodesics \eqref{gammapm} which intersect with $\mathcal{E}$ at the endpoints of $\mathcal{E}_{reg}$ (the solid blue line).
 \label{intrinsic} }
\end{figure}

Following the above prescription, the endpoints $\partial_\pm \mathcal{A}$ are pushed to the following positions
\begin{align}\label{regulatedpA}
 \partial_\pm \mathcal{A}:~~(\pm\frac{l_u}{2}\mp \frac{1}{2 r_I},\pm\frac{l_v}{2},r_I)\,.
 \end{align}
Note that all the null geodesics $\bar{\mathcal{L}}_{\bar{v}_0}$ lie on $\mathcal{N}_{\pm}$ are given in \eqref{nullgeodesics} and $\bar{v}_0$ denotes the $v$ coordinate of the points where $\bar{\mathcal{L}}_{\bar{v}_0}$ intersect with $\mathcal{E}$. It is easy to find that the two null geodesics $\gamma_{\pm}$ emanating from $\partial_{\pm}\mathcal{A}$ \eqref{regulatedpA} are just given by $\bar{\mathcal{L}}_{\bar{v}_0}$ with $\bar{v}_0=\pm (\frac{l_v}{2}+\frac{1}{2}\log(l_u r_I))$, i.e.
\begin{align}\label{gammapm}
\gamma_\pm:~~u=\mp\frac{1}{2} \left(\frac{1}{r}-l_u\right),\qquad v=\pm\left(\frac{l_v}{2}+\frac{1}{2}\log(l_u r_I)\right)\mp\frac{1}{2} \log (l_u r)\,. 
\end{align}
When $r=\frac{1}{l_u}$ we get the end points of $\mathcal{E}_{reg}$. See Fig. \ref{intrinsic} for an overall picture of our construction. The length of the regulated curve $\mathcal{E}_{reg}$ is just
\begin{align}
Length(\mathcal{E}_{reg})=l_v+\log(l_ur_I)\,,
\end{align}
which reproduces the result \eqref{SeeLM} with the UV/IR relation \eqref{UVIR}. In appendix \ref{appendixB}, we show that, similar to the flat case \cite{Jiang:2017ecm}, the $\mathcal{E}_{reg}$ is the saddle among all the geodesics that connect $\gamma_\pm$ \eqref{gammapm}.

\section{Towards the generalized gravitational entropy for spacetimes with non-Lorentz duals}\label{discussion}
Based on the above discussions, we are ready to generalize our intrinsic prescription to calculate the generalized gravitational entropy for general spacetimes with non-Lorentz invariant duals. As expected the prescription only evolves extremal surfaces $\mathcal{E}$ and their associated normal null hypersurfaces $\mathcal{N}_{\pm}$, which are available when the bulk metric is given. 

For a bulk spacetime $\mathcal{M}$ and its asymptotic boundary $\mathcal{B}$ at $r\to\infty$, when a holography is conjectured between the bulk gravity theory (here we only consider Einstein gravity) and the field theory on $\mathcal{B}$. The prescription is in the following

\begin{enumerate} 

\item Firstly we should figure out the causal structure of the boundary field theory, using either the boundary null geodesics (or hypersurfaces) when the metric background is fixed, or the Rindler method. In other words, for a given subregion $\mathcal{A}$ we should find out the corresponding causal development $\mathcal{D}_{\mathcal{A}}$. For non-Lorentzian field theories, the causal developments usually looks like a strip (or a solid cylinder) rather than a diamond.
 
\item For a general spacelike extremal surface we study its normal null hypersurfaces $\mathcal{N}_{\pm}$. As we have discussed before, the $\mathcal{E}$ can be determined by following requirement{\footnote{For spacetimes with relativistic duals, this requirement naturally lead to the requirement that $\mathcal{E}$ should be anchored on $\partial\mathcal{A}$. }}
\begin{align}
\mathcal{E}:~\mathcal{N}_{\pm}\cap\mathcal{B} \supset \partial\mathcal{D}_{\mathcal{A}}\,,
\end{align}
which is just the requirement for the consistency between the bulk and boundary causal structures.  

\item Then we push the dual field theory to the cutoff boundary at some large radius $r=r_I$ {\footnote{The cutoff radius $r_I$ should be properly related to the UV cutoff of the field theory, and the relation should be discussed case by case.}}.  During we push the boundary, it is crucial how the entangling surface $\partial\mathcal{A}$ moves. The main requirement is that $\partial \mathcal{A}$ should adapt to the consistency of the bulk and boundary causal structures on $r=r_I${\footnote{For example, in the relativistic holography cases where $\mathcal{E}$ is anchored on $\partial\mathcal{A}$, the only way to satisfy this requirement is to push $\partial\mathcal{A}$ along $\mathcal{N}_+\cup\mathcal{N}_-=\mathcal{E}$.}}. For the cases with non-Lorentz invariant duals, we should keep $\partial\mathcal{A}$ on $\mathcal{N}_\pm$ and keep the coordinates, whose UV cutoff can be taken to be zero, fixed.

\item On $\mathcal{N}_\pm$, there are null geodesics $\gamma_{\pm}$ (or codimension two null hypersufaces) emanating from $\partial\mathcal{A}$ at the cutoff boundary $r=r_{I}$. We cut off the extremal surface $\mathcal{E}$ at the place where $\gamma_{\pm}$ intersect with $\mathcal{E}$. Then we get the regulated extremal surface $\mathcal{E}_{reg}$ and the regulated holographic entanglement entropy
\begin{align}
S_{\mathcal{A}}=\frac{Area\left(\mathcal{E}_{reg}\right)}{4 G}\,.
\end{align}

\end{enumerate}
The first two steps show how to use the consistency of the causual structures to determine the $\mathcal{E}$ corresponding to the boundary subregion $\mathcal{A}$ in question, while the last two steps tell us how to regulate $\mathcal{E}$ using the null geodesics on $\mathcal{N}_\pm $. We conjuecture that way we push $\partial \mathcal{A}$ will get it to the right modular plane that determines the regulation. 

For the cases with locally defined modular Hamiltonian, the fine structure analysis with the modular planes gives a strong support for the validity of our prescription. However in more general cases, the modular Hamiltonian is usually non-local, so the modular planes can no longer be described as certain bulk codimension one surfaces. Instead they should be defined in a more abstract way. 
We would like to stress that, for the more general cases our prescription is still applicable. Though in general the modular Hamiltonian is non-local, effectively it can be locally defined in some special bulk and boundary regions: the region near the $\mathcal{N}_{\pm}$ and the region near $\partial \mathcal{D}_{\mathcal{A}}$. This indicates the modular planes can also be locally defined in these regions and intersect with $\mathcal{N}_{\pm}$ on the null geodesics that form $\mathcal{N}_\pm$. Note that our prescription is insensitive to how the modular planes look like outside these regions, so with all the geometric quantities (including the cutoff points of $\mathcal{A}$, the extremal surface $\mathcal{E}$ and the null geodesics or hypersurfaces $\gamma_\pm$ on $\mathcal{N}_{\pm}$) we need to apply our prescription inside these regions, our prescription is still applicable. We conjecture that the result we get by applying our prescription is still the right holographic entanglement entropy. 

In the following we give an argument for this statement. When we need to do the regulation, we can divide the $\mathcal{A}$ and $\mathcal{E}$ into two parts
\begin{align}
\mathcal{A}=\mathcal{A}_{cut}\cup\mathcal{A}_{reg}\,,\qquad \mathcal{E}=\mathcal{E}_{cut}\cup\mathcal{E}_{reg}.
\end{align}
Here $\mathcal{A}_{cut}$ is the infinitesimal part of $\mathcal{A}$ we cut off. Also the total entanglement entropy can be divided into two parts which are contributed from $\mathcal{A}_{cut}$ and $\mathcal{A}_{reg}$ respectively
\begin{align}\label{totalentanglement}
S_{\mathcal{A}}=s_{\mathcal{A}}({\mathcal{A}_{cut}})+ s_{\mathcal{A}}({\mathcal{A}_{reg}})=\frac{Area(\mathcal{E})}{4G}\,.
 \end{align} 
Though the fine correspondence for the points on $\mathcal{A}_{reg}$ no longer exists when the modular Hamiltonian is non-local, it still holds between the points on $\mathcal{A}_{cut}$ and the points on $\mathcal{E}_{cut}$. This is because $\mathcal{A}_{cut}$ is in the near $\partial\mathcal{D}$ region, where the modular flow effectively has a local description. So according to the fine correspondence $\mathcal{A}_{cut}$ correspond to some part of $\mathcal{E}$ which we call  $\mathcal{E}_{cut}$, such that
\begin{align}
s_{\mathcal{A}}({\mathcal{A}_{cut}})=\frac{Area(\mathcal{E}_{cut})}{4G}\,.
\end{align}
Together with \eqref{totalentanglement}, we find the regulated entanglement entropy is given by
\begin{align}
S_{\mathcal{A}}^{reg}= s_{\mathcal{A}}({\mathcal{A}_{reg}})=\frac{Area(\mathcal{E}_{reg})}{4G}\,.
\end{align}

\section{Generalised gravitational entropy for 3-dimensional flat space}\label{flat}
In this section we apply our prescription to the case of 3-dimensional flat holography \cite{Barnich:2010eb,Bagchi:2010eg,Bagchi:2012cy}. In this holography, the 3-dimensional asymptotic flat spacetime is conjectured to be dual to a field theory invariant under the BMS$_3$ group (BMSFT). The BMS$_3$ group is the asymptotic symmetry group of flat space enhanced from the Poincar\'e group, and the BMSFT can be considered as the ultra-relativistic limit of a CFT$_2$. The holographic calculation, as well as the geometric picture, of the entanglement entropy for BMSFTs are given in \cite{Jiang:2017ecm} with the Rindler method. We will show that our prescription can easily reproduce the results in \cite{Jiang:2017ecm} without the Rindler transformations and modular flows.

In particular we consider the following classical solutions of Einstein gravity with vanishing cosmological constant in Bondi gauge
\begin{align}\label{fbtz}
ds^2= M du^2-2 dudr+ J du d\phi+r^2 d\phi^2.
\end{align}
The above solutions are usually classified into three types:
\begin{itemize}
\item  $M=-1,J=0$: Global Minkowski, which duals to the zero temperature BMSFT on the cylinder with $\phi\sim\phi+2\pi$.

\item   $M=J=0$: Null-orbifold, with $\phi$ decompactified this duals to the zero temperature BMSFT on the plane.
\item  $M>0 $: Flat Space Cosmological solutions (FSC), which duals to BMSFT at finite temperature.
\end{itemize}

The asymptotic boundary (null infinity) $\mathcal{B}$ settles at $r = r_I \to\infty$ with a fixed background metric
\begin{align}
ds^2=0 du^2+d\phi^2\,,
\end{align}
which is degenerate. On $\mathcal{B}$ the null direction is characterized by $u$. The subregion we study is a single interval
\begin{align}
\mathcal{A}:\qquad (  - l_u/2 ,  - l_\phi/2  ) \rightarrow (  l_u/2 , l_\phi/2 )\,.
\end{align}
Since $u$ is the null direction, the domain of causality $\mathcal{D}_{\mathcal{A}}$ is just a strip along the $u$ direction
\begin{align}
\mathcal{D}_{\mathcal{A}}:\qquad \{-\frac{l_\phi}{2}\leq \phi \leq \frac{l_\phi}{2}\}\,.
\end{align}
Asymptotically we should have
\begin{align}\label{DA}
\partial\mathcal{D}_{\mathcal{A}}:\qquad \phi=\pm\frac{l_\phi}{2}+\mathcal{O}\left(\frac{1}{r_I}\right)\,,\qquad r=r_I\,.
\end{align}

It will be quite subtle to apply our prescription in the Bondi gauge, so we apply it in the Cartesian coordinates. Another advantage of using the Cartesian coordinates is that we do not need to solve the Einstein equations, because the geodesics and their null normal hypersurfaces are just straight lines and null planes.

\subsection{Null-orbifold}
We choose the coordinate transformation between the Null-orbifold and the Cartesian coordinates to be
\begin{align}
t &={l_\phi\over4}r +{2\over l_\phi}(u+\frac{r\phi^2}{2})  \,,\\
x &= ({l_u\over l_\phi}+r\phi)  \,,\\
y &={l_\phi\over4} r-{2\over l_\phi}(u+\frac{r\phi^2}{2})\,.
\end{align}
Here we have adjusted the transformation in advance by some proper Poincar\'e transformation and the coefficients are chosen to be related to the parameters $l_u$ and $l_\phi$ which characterize the boundary interval $\mathcal{A}$, hence the curve $\mathcal{E}$ can be characterized by a single coordinate $y$ and settled at $t=x=0$. Of course one can begin with free coefficients and settle them down one by one through the matching condition \eqref{causalconsistence}. It is easy to check that, up to a Poincar\'e transformation for the Cartesian coordinates, we have
\begin{align}
ds^2&= -2 dudr +r^2 d\phi^2= -dt^2+dx^2+dy^2\,.
\end{align}

Then in Cartesian coordinates $\partial\mathcal{D}_{\mathcal{A}}$ \eqref{DA} and the two endpoints $\partial_\pm\mathcal{A}$ of $\mathcal{A}$ are given by
\begin{align}\label{flatpda}
\partial\mathcal{D}_{\mathcal{A}}:& \quad \left\{t,x,y\right\}=\left\{\frac{l_\phi}{2}r_I+\mathcal{O}\left(r_I^{0}\right),\pm \frac{l_\phi}{2} r_I+\mathcal{O}\left(r_I^{0}\right), -\frac{2 u}{l_\phi}\right\}\,,
\\
\partial_\pm\mathcal{A}:& \quad \left\{t,x,y\right\}=\left\{\frac{l_\phi}{2}r_I \pm\frac{l_u}{l_\phi},\pm \frac{l_\phi}{2} r_I+\frac{l_u}{l_\phi}, \mp\frac{l_u}{l_\phi}\right\}\,.
\end{align}
It is easy to see that the spacelike geodesic $\mathcal{E}$ and the associated $\mathcal{N}_\pm$ that asymptotically satisfy \eqref{causalconsistence} are just given by 
\begin{align}\label{flatE}
\mathcal{E}:&\quad \left\{x=0,\quad t=0\right\}\,,
\\\label{flatN}
\mathcal{N}_\pm:&\quad \left\{x=\pm t\right\}\,.
\end{align}
Obviously $\mathcal{N}_\pm$ \eqref{flatN} will asymptotically go through $\partial \mathcal{D}_{\mathcal{A}}$ \eqref{flatpda}. 

\begin{figure}[h] 
   \centering
        \includegraphics[width=0.5\textwidth]{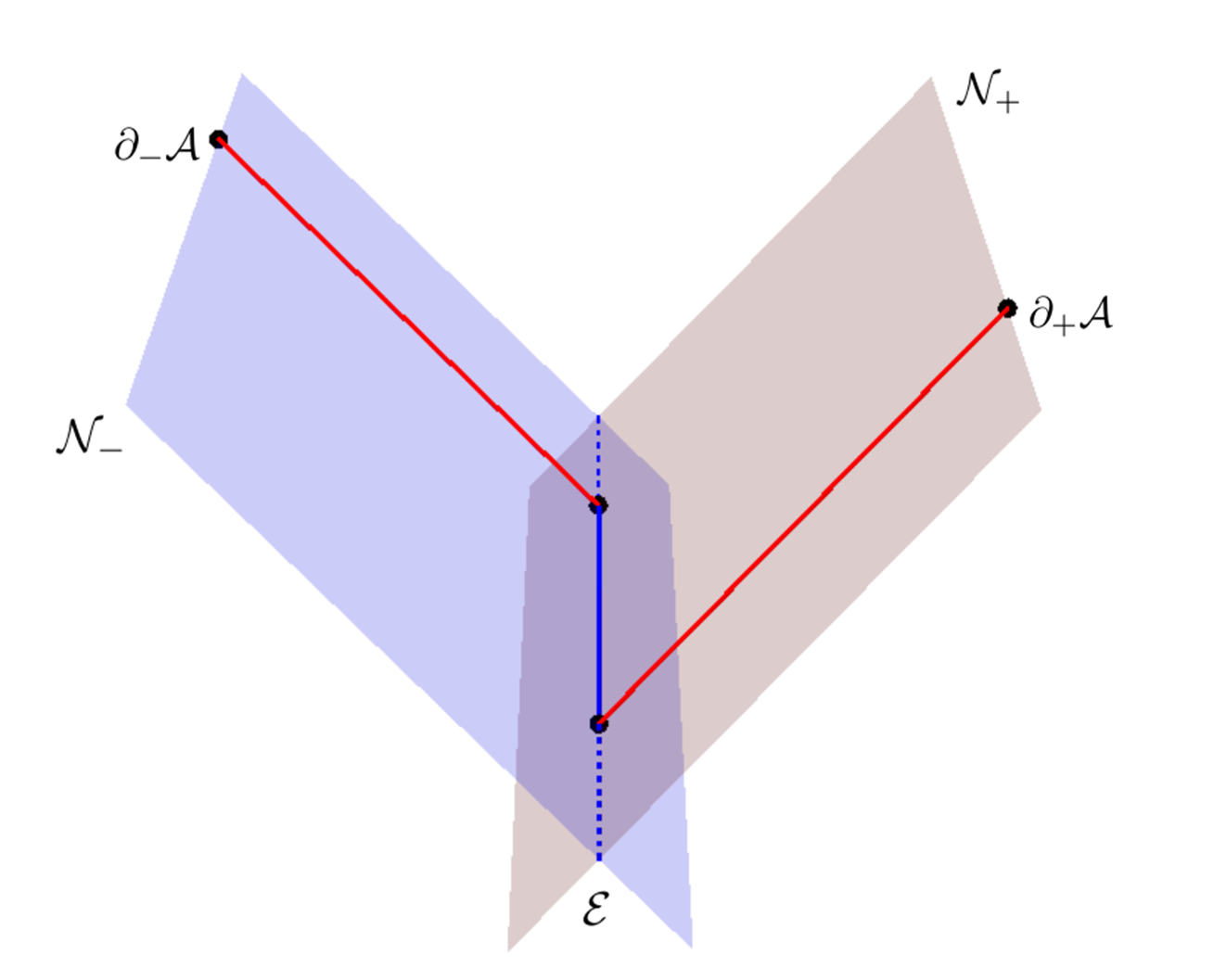}  
 \caption{The two red lines are the null rays $\gamma_\pm$ emanating from $\partial_\pm\mathcal{A}$ and normal to $\mathcal{E}$. The geodesic $\mathcal{E}$ is cut off where it intersect with $\gamma_\pm$. The solid blue segment is just our $\mathcal{E}_{reg}$.
 \label{intrinsicflat} }
\end{figure}

Then we regulate $\mathcal{E}$ using null geodesics on $\mathcal{N}_\pm$. The two null geodesics $\gamma_\pm$ on $\mathcal{N}_\pm$ that emanating from $\partial_\pm\mathcal{A}$ are given by
\begin{align}
\gamma_\pm:\quad \left\{x=\pm t\,,\quad y=\mp\frac{l_u}{l_\phi}\right\}\,.
\end{align}
Note that in this case there is no need to introduce UV cutoffs in the $u$ and $\phi$ direction since they can be taken to be zero without introducing divergence to the entanglement entropy{\footnote{However when we consider gravity with gravitational anomaly, for example the topological massive gravity \cite{Deser:1981wh,Deser:1982vy}, a divergent contribution will arise due to the anomaly \cite{Jiang:2017ecm}.}}. So when we push the boundary, we keep the $u$ sand $\phi$ coordinate of $\partial_\pm\mathcal{A}$ fixed. This means $\partial_\pm\mathcal{A}$ moves along $\gamma_\pm$ and the $\mathcal{E}_{reg}$ will be independent of the choice of $r_I$. The overall picture of our construction is shown in Fig. \ref{intrinsicflat}.

According to our prescription, the points where $\gamma_\pm$ intersect with $\mathcal{E}$ are the place we cut $\mathcal{E}$ off. Then we get
\begin{align}
\mathcal{E}_{reg}:\quad \left\{x=t=0\,,\quad -\frac{l_u}{l_\phi}\leq y \leq \frac{l_u}{l_\phi}\right\}\,.
\end{align}
Accordingly we have
\begin{align}
S_{\mathcal{A}}=\frac{Area\left(\mathcal{E}_{reg}\right)}{4G}=\frac{1}{2G}\frac{l_u}{l_\phi}\,,
\end{align}
which reproduces the results in \cite{Bagchi:2014iea,Basu:2015evh,Jiang:2017ecm,Hijano:2017eii}.

\subsection{Global Minkowski}
The global Minkowski space
\begin{align}
ds^2=-du^2-2 du dr+r^2 d\phi^2\,,
\end{align}
can be transformed to the Cartesian coordinates by the following transformation
\begin{align}
t=&(r+u)\csc\frac{l_\phi}{2}+\frac{r\cos \phi}{2}\left(\tan \frac{l_\phi}{4}-\cot \frac{l_\phi}{4}\right)\,,
\cr
x=&r \sin\phi+\frac{1}{2}l_u\csc\frac{l_\phi}{2}\,,
\cr
y=&r \cos\phi \csc \frac{l_\phi}{2}-(r+u)\cot \frac{l_\phi}{2}\,.
\end{align}

Then in Cartesian coordinates we have
\begin{align}
\partial\mathcal{D}_{\mathcal{A}}:& \quad \left\{t,x,y\right\}=\left\{\sin\frac{l_\phi}{2}r_I+\mathcal{O}\left(r_I^{0}\right),\pm \sin\frac{l_\phi}{2} r_I+\mathcal{O}\left(r_I^{0}\right), -u \cot\frac{l_\phi}{2}\right\}
\\
\partial_\pm\mathcal{A}:& \quad \left\{t,x,y\right\}=\left\{\frac{\csc\frac{l_\phi}{2}}{2}\left(l_u\pm 2 r_I\sin^2\frac{l_\phi}{2}\right),\pm r_I\sin\frac{l_\phi}{2},\mp \frac{l_u\cot \frac{l_\phi}{2}}{2}\right\}
\end{align}
Similarly we get the spacelike geodesic $\mathcal{E}$ and the associated $\mathcal{N}_\pm$ that asymptotically satisfy the requirement \eqref{causalconsistence}
\begin{align}
\mathcal{E}:&\quad \left\{x=0,\quad t=0\right\}\,,
\\
\mathcal{N}_\pm:&\quad \left\{x=\pm t\right\}\,.
\end{align}
The two null geodesics $\gamma_\pm$ on $\mathcal{N}_\pm$ that emanating from $\partial_\pm\mathcal{A}$ are just given by
\begin{align}
\gamma_\pm:\quad \left\{x=\pm t\,,\quad y=\mp l_u\cot \frac{l_\phi}{2}\right\}\,.
\end{align}
Quite straightforwardly we get
\begin{align}
S_{\mathcal{A}}=\frac{Area\left(\mathcal{E}_{reg}\right)}{4G}=\frac{l_u}{2G}\cot \frac{l_\phi}{2}\,.
\end{align}

\subsection{Flat Space Cosmological solutions}
The coordinate transformations from FSC to the Minkowski space can be given by
\begin{align}
  r &= \sqrt{M(t'^2-x'^2)+r_c^2 } \,,
  \\
  \phi &=-\frac{1}{ \sqrt{  M}} \log\frac{\sqrt{M}(t'-x')}{r+\frac{J}{2\sqrt{M}}}\,,
   \\
  u &= \frac{1}{ M}\Big(   r -\sqrt{M}y'-\frac{J}{2} \phi \Big)\,.
\end{align}
The above transformations show that the FSC can be considered as a quotient of the Minkowski space, because the region outside the cosmological horizon $r_c=\frac{J}{2\sqrt{M}}$ only covers a quarter of the Minkowski space $t'\geq |x'|$.

As in the previous two cases, we can apply an additional Poincar\'e transformation to a new set of Cartesian coordinates $\{t,x,y\}$ thus the corresponding $\mathcal{E}$ and $\mathcal{N}_\pm$ are just given by \eqref{flatE} and \eqref{flatN}. Under this requirement $\partial\mathcal{D}_{\mathcal{A}}$ \eqref{DA} should asymptotically satisfy $t=\pm x$, then we find that the additional Poincar\'e transformation is just
\begin{align}
t=&t' \cosh \eta-y'\sinh \eta\,,
\\
x=&x'+s_0\,,
\\
y=&y' \cosh \eta-t'\sinh\eta\,,
\end{align}
where
\begin{align}
\eta=\text{arccosh}\left[\text{coth}\frac{l_\phi\sqrt{M}}{2}\right]\,,\qquad s_0=\frac{(J l_\phi+2 l_u M)\text{csch}\frac{l_\phi \sqrt{M}}{2}}{4 \sqrt{M}}\,.
\end{align}
Then in Cartesian coordinates we have
\begin{align}
\partial\mathcal{D}_{\mathcal{A}}:& \quad \left\{t,x,y\right\}=\Big\{\frac{\sinh\frac{l_\phi \sqrt{M}}{2}}{\sqrt{M}} r_I+\mathcal{O}\left(r_I^{0}\right),\pm \frac{\sinh\frac{l_\phi \sqrt{M}}{2}}{\sqrt{M}} r_I+\mathcal{O}\left(r_I^{0}\right),
\cr
&~~~~~~~~~~~~~~~~~~~~~\frac{\pm 2 J\mp \sqrt{M}\left(J l_\phi\pm 4 M u\right)\coth \frac{l_\phi \sqrt{M}}{2}}{4M}\Big\}\,,
\\
\partial_\pm\mathcal{A}:& \quad y=\frac{\pm 2 J\mp \sqrt{M} \coth \left(\frac{\sqrt{M} l_{\phi }}{2}\right) \left(J l_{\phi }\pm 2 M l_u\right)}{4 M}\,.
\end{align}
For simplicity we only list the $y$ coordinates of $\partial_\pm \mathcal{A}$.
The two null geodesics $\gamma_\pm$ emanating from $\partial_\pm\mathcal{A}$ are given by
\begin{align}
\gamma_\pm:\quad \Big\{x=\pm t\,,\quad y=\frac{\pm 2 J\mp \sqrt{M} \coth \left(\frac{\sqrt{M} l_{\phi }}{2}\right) \left(J l_{\phi }\pm 2 M l_u\right)}{4 M}\Big\}\,.
\end{align}
Again we reproduce the right result
\begin{align}
S_{\mathcal{A}}=\frac{Area\left(\mathcal{E}_{reg}\right)}{4G}=\frac{1}{4G}\left|\frac{\left(J l_{\phi }+2 M l_u\right)}{2 \sqrt{M}}\coth \left(\frac{\sqrt{M} l_{\phi }}{2}\right) -\frac{J}{M}\right|\,.
\end{align}

\section{Discussion}
In this paper we demonstrate how the RT formula fails to give the geometric picture of the holographic entanglement entropy for spacetimes with non-Lorentz invariant duals. We extend the discussion of \cite{Lewkowycz:2013nqa} to holographies beyond AdS/CFT. The two main points which are crucial for this extension include the requirement for the consistency between the bulk and boundary causal structures, and the introduction of null geodesics (or hypersurfaces) to control the regulation of entanglement entropy. Since $\gamma_\pm$ are null thus do not contribute to the total length, one may consider a new $\mathcal{E}=\mathcal{E}_{reg}\cup\gamma_+\cup\gamma_-$ such that $\mathcal{E}$ is anchored on $\partial\mathcal{A}$ as the RT formula. However we do not suggest to do that. First of all, $\mathcal{E}_{reg}$ is fixed under the modular flow (or replica symmetry) while $\gamma_\pm$ are not. So they play totally different roles in the replica story. Also they play different roles in the new  extrapolate dictionary \eqref{newdictionary} which we will discuss later. Secondly, the requirement that the new $\mathcal{E}$ is anchored on $\partial\mathcal{A}$ does not help to determine the geometric picture as the RT formula, because usually the null geodesics emanating from $\partial\mathcal{A}$ are not unique. So we still need the help of \eqref{causalconsistence} to determine $\mathcal{E}$. At last, unlike $\mathcal{E}$, the null geodesics $\gamma_\pm$ depend on the cutoff $\epsilon$.

Since the RT formula stimulates numerous insights in our understanding of the AdS/CFT in many aspects, we expect the parallel discussions based on our prescription can help us to better understand the holographies beyond AdS/CFT. In the following we list a few interesting problems that may be solved based on our new geometric picture.

\subsection*{Holographic entanglement entropy for Lifshitz spacetime}
One particular interesting class of non-Lorentzian field theories are those with Lifshitz symmetries. The dual spacetime, which we call Lifshitz spacetimes, was proposed in \cite{Kachru:2008yh,Taylor:2015glc}. It has been shown in \cite{Gentle:2015cfp} that the normal null hypersurfaces emanating from the RT (or HRT) surface in Lifshitz spacetime can not reach the boundary and thus fail to satisfy \eqref{causalconsistence}. This implies the inconsistency between the bulk and boundary casual structures, so the RT formula should fail in this case according to our discussion. This means the calculations of the holographic entanglement entropies for Lifshitz-type theories \cite{lif1,lif2,lif3,lif4,lif6,lif7,lif8,lif9} following the RT formula should not be correct. Also they are not consistent with the recent numerical results \cite{Nlif1,Nlif2,Nlif3,Nlif4} of entanglement entropies for free Lifshitz-type theories\footnote{However, the Lifshitz-type field theories which have gravity dual are supposed to be strongly coupled and in the large $N$ limit. Unlike the case of 2-d CFT, WCFT and BMSFT where there are much more symmetries, the formula of the two point function (or entanglement entropy) cannot be determined by symmetries in Lifshitz-type field theories. So the comparison with these numerical results may not make much sence.}. 

Assuming the gravity dual is an Einstein gravity, our prescription can be applied to the Lifshitz spacetimes and give new holographic predictions for the entanglement entropy of holographic Lifshitz-type theories. Also it is argued in \cite{Griffin:2012qx} that Horava-Lifshitz gravity is the minimal holographic dual for Lifshitz field theories (see also \cite{Cheyne:2017bis} along this line). In this case our prescription need corrections.

\subsection*{Holographic entanglement entropy in higher dimensional flat space}
The geometric picture for holographic entanglement entropy in 3-dimensional flat holography is carried out in \cite{Jiang:2017ecm} by Rindler method. It is then straightforward to ask whether we can extend the calculation to the case of flat holography in 4-dimensions, which has recently attracted lots of attention (see \cite{Strominger:2017zoo} for a recent review). Unfortunately the Rindler method get much more complicated in higher dimensions. Since the prescription proposed in this paper is intrinsic and has natural extension to higher dimensions, it will be more promising to solve this problem by our prescription.

\subsection*{New extrapolate dictionary between boundary operators and bulk matter fields}
Following the replica trick, the entanglement entropy in a field theory can be computed by evaluating the two point function of twist operators located at the boundary endpoints. So the new geometric picture of entanglement entropy with extra null geodesics gives a new holographic description for the two point correlation functions. Motivated by this picture and a similar prescription \cite{Hijano:2015rla,Castro:2017hpx} for calculating holographic conformal blocks in the probe limit in AdS/CFT, the authors of \cite{Hijano:2017eii} calculated the Poincar\'e blocks (or global BMS blocks) for BMSFT holographically by extremizing the length of a network of geodesics connected to the operators at the boundary through certain null geodesics. And the results match with the calculations \cite{Bagchi:2017cpu} on the field theory side.

Based on the above results, an extrapolate dictionary is proposed in \cite{Hijano:2017eii} (see also \cite{Hijano:2018nhq}) for flat holography. More explicitly, to each boundary point $x$ where we inject an operator, we attach a null line $\gamma_x$, which is similar to the null lines $\gamma_\pm$ in our geometric picture for entanglement entropy. The proposed extrapolate dictionary is then to attach a position space Feynman diagram to the null lines and integrate the position of the legs over an affine parameter
along $\gamma_x$
\begin{align}\label{newdictionary}
\left<\mathcal{O}(x_1)\mathcal{O}(x_2)\cdots \right>=\int_{\gamma_{x_1}}d\lambda_1\int_{\gamma_{x_2}}d\lambda_2 \cdots \left<\psi(\lambda_1)\psi(\lambda_2)\cdots\right>\,.
\end{align}
Here $\psi$ denotes the bulk matter fields that correspond to the boundary operators $\mathcal{O}$, and $\lambda_i$ parametrizes the null geodesic $\gamma_{x_i}$ emanating from the boundary point $x_i$. The above extrapolate dictionary in flat holography is totally different from the one \cite{Witten:1998qj} in AdS/CFT, and gives another entry to study flat holography.

Our analysis shows that the geometric picture for holographic entanglement entropy in spacetimes with non-Lorentzian duals should in general include the extra null geodesics (or null hypersurfaces). So in these cases, the right extrapolate dictionary for boundary operators and bulk matter fields should be similar to \eqref{newdictionary} rather the one \cite{Witten:1998qj} in AdS/CFT. It will be very interesting to test the dictionary \eqref{newdictionary} and calculate the correlation functions holographically in other spacetimes with non-Lorentz invariant duals.

\subsection*{Acknowledgments} 
We would like to thank Wei Song and Rong-xin Miao for initial collaboration on this project and insightful discussions. We thank Matthew Headrick and Hongliang Jiang for helpful discussions. We thank Wei Song, Chang-pu Sun, Pin Yu, Wenbin Yan and the Yau Mathematical Sciences Center of Tsinghua University for support. We would also like to thank the Center of Mathematical Sciences and Applications at Harvard University for hospitality during the development of this work. This work is supported by NSFC Grant No.11805109.

\appendix

\section{Spacelike geodesics in AdS$_3$}\label{geodesics}
The spacelike geodesics in the AdS$_3$ \eqref{btz} with $T_u=0$ satisfy the following equations of motion
\begin{align}
\label{eq31}
\frac{c_1}{\ell^2}&=r \dot{v}\,,
\\
\label{eq32}
\frac{c_2}{\ell^2}&=r \dot{u}+T_v^2 \dot{v}\,,
\\
\label{eq33}
\frac{1}{\ell^2}&=T_v^2\dot{v}^2+2 r \dot{u}\dot{v}+\frac{\dot{r}^2}{4r^2}\,,
\end{align}
where we $c_{1}$ and $c_2$ are two integration constants, satisfying $c_1c_2>0$, and dot represent differential with respect to the affine parameter $s$.
From (\ref{eq31}) and (\ref{eq32}) we get
\begin{align}
\dot{u}={c_2 r-c_1 T_v^2\over \ell^2 r^2}\,,
\qquad
\dot{v}={c_1\over r\ell^2}\,.
\end{align}
Substituting the above equations into (\ref{eq33}) we get a radial equation 
\begin{align}
\dot{r}=\pm \frac{2 \sqrt{r \left(\ell ^2 r-2 c_1 c_2\right)+T_v^2 c_1^2}}{\ell}\,,\label{rdot}
\end{align}

We get three types of spacelike geodesics by adjusting the value of $|c_2|$. Firstly when
\begin{align}\label{rstar}
|c_2|>\ell T_v\,,
\end{align}
we find $\dot{r}=0$ at 
\begin{align}
r_{\pm}=\frac{c_1 c_2\pm\sqrt{c_1^2 \left(c_2^2-\ell ^2 T_v^2\right)}}{\ell ^2}>0\,.
\end{align}
This type of geodesic is anchored on the boundary and has a turning points at $r_+$, thus belongs to the RT curves. The second type of geodesic satisfies $|c_2|<\ell T_v$ thus has no turning points. 

Both of the above two types do not satisfy our requirement \eqref{causalconsistence}. The third type of geodesic that never touches the boundary arises when
\begin{align}\label{c2S}
|c_2|=\ell T_v\,.
\end{align}
In this case we find that,
at $r={c_1T_v\over\ell},$
\begin{align}
\dot{r}=\dot{u}=0,\qquad \dot{v}={1\over c_2}=\pm{1\over\ell T_v}\,.
\end{align} 
The corresponding geodesics $\mathcal{E}$ lies along the $v$ direction at a fixed radius
\begin{align}\label{S}
\mathcal{E}:~~\{r=\frac{c_1 T_v}{\ell}\,,\quad u=u_0\}\,,
\end{align}
where $u_0$ is an arbitrary constant. In the case \eqref{tu0tv1} we study, we set $T_v=1$. Note that, for our coordinates \eqref{btz}, the global AdS correspond to $T_u=T_v=\frac{i}{2}$ and Poincar\'{e} AdS correspond to $T_u=T_v=0$. In these two cases there are no spacelike geodesics that do not touch the boundary. For the given interval \eqref{interval}, the $\mathcal{E}$ that satisfies the requirement \eqref{causalconsistence} is just \eqref{S} with $c_1=\frac{1}{l_u}$.

\section{Entanglement contour for WCFT}\label{appendixA}
\subsection{Entanglement contour from the fine structure}
The entanglement contour function is a density function of entanglement. In other words it describes the distribution of the contribution to the total entanglement entropy from each point of $\mathcal{A}$
\begin{align}
S_{\mathcal{A}}=\int_{\mathcal{A}}s_{\mathcal{A}}(v)dv\,.
\end{align}
Here we parametrize $\mathcal{A}$ with the $v$ coordinate.
The authors of \cite{Vidal} proposed a set of requirements for the contour functions{\footnote{However the complete list of requirements that uniquely determines the contour is still not available.}}. Few analysis of the contour functions for bipartite entanglement have been explored in \cite{Botero,Vidal,PhysRevB.92.115129,Coser:2017dtb,Tonni:2017jom}. Also its fundamental definition is still not established.

In the previous section we propose that $Length(\mathcal{E}_{2})$ captures the contribution from $\mathcal{A}_2$ to the entanglement entropy $S_\mathcal{A}$. In other words our fine structure analysis gives a holographic interpretation for the contour function. Consider $\mathcal{A}$ to be a straight line  \eqref{straightA}, according to the fine correspondence \eqref{v0vp} we get the contour function $s_{\mathcal{A}}(v)$ for $S_\mathcal{A}$
\begin{align}\label{contour}
s_{\mathcal{A}}(v)=\frac{1}{4G}\left(1+\frac{2 l_v}{l_v^2-4 v^2}\right)\,.
\end{align}

Let us define 
\begin{align}\label{SA22}
s_{\mathcal{A}}(\mathcal{A}_2)=\int_{\mathcal{A}_2}s_{\mathcal{A}}(v)dv=\frac{Length(\mathcal{E}_2)}{4G}\,.
\end{align}
According to the fine correspondence \eqref{v0vp} or the contour function \eqref{contour}, we have
\begin{align}\label{SA2}
s_{\mathcal{A}}(\mathcal{A}_2)=\frac{1}{4G}\left(v'_2-v'_1+\text{arctanh}\frac{2 v'_2}{l_v}-\text{arctanh}\frac{2 v'_1}{l_v}\right)
\end{align}
where $v'_1$ and $v'_2$ are the $v$ coordinates of the two endpoints of $\mathcal{A}_2$. 

\begin{figure}[h] 
   \centering
        \includegraphics[width=0.4\textwidth]{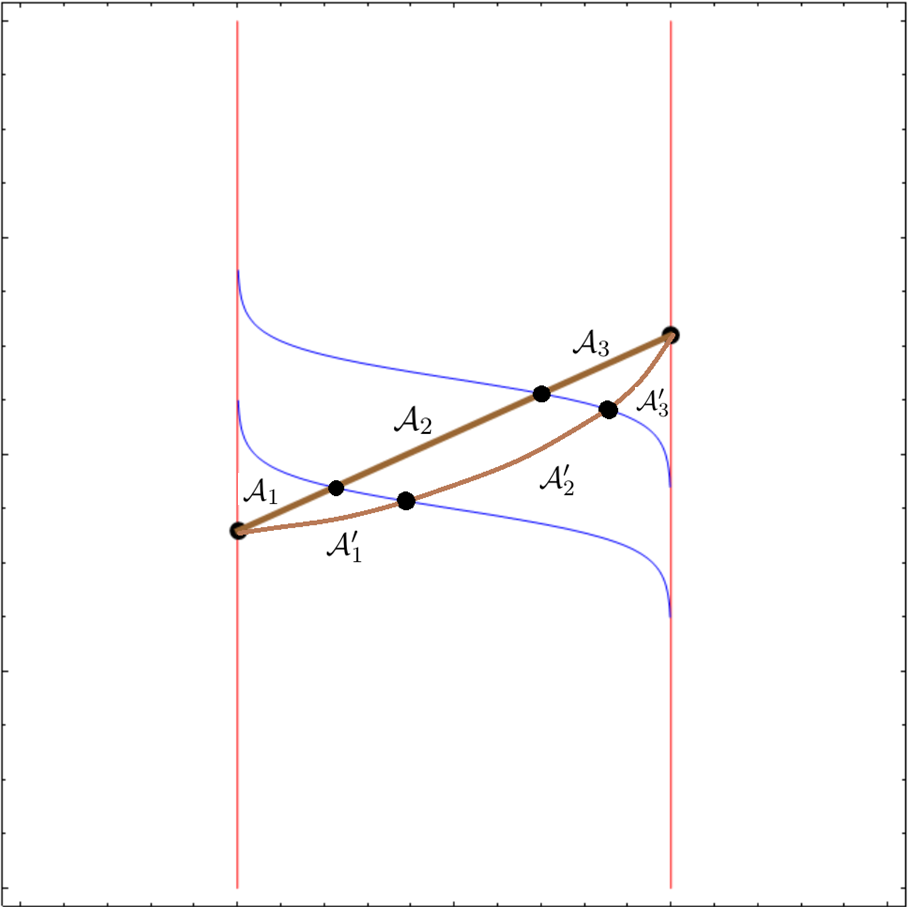}  
 \caption{ The two blue lines are two modular flow lines $\mathcal{L}_{v_1}$ and $\mathcal{L}_{v_2}$ on the boundary. Here $\mathcal{A}_2$ and $\mathcal{A}'_2$ intersect with the same bunch of modular planes between $\mathcal{P}(v_1)$ and $\mathcal{P}(v_2)$.
\label{A2} }
\end{figure}

Furthermore we consider two intervals $\mathcal{A}$ and $\mathcal{A}'$ that have the same causal development $\mathcal{D}_{\mathcal{A}}=\mathcal{D}_{\mathcal{A}'}$. The two arbitrary  boundary modular flow lines $\mathcal{L}_{v_1}$ and $\mathcal{L}_{v_2}$ that divide $\mathcal{A}$ ($\mathcal{A}'$) into three part $\mathcal{A}_1,\mathcal{A}_2$ and $\mathcal{A}_3$ ($\mathcal{A}'_1,\mathcal{A}'_2$ and $\mathcal{A}'_3$). See Fig.\ref{A2}. According to our prescription, any $\mathcal{A}_2$ that goes through the same bunch of modular planes should correspond to the same $\mathcal{E}_2$ on $\mathcal{E}$. Then we should have the following causal property for $S_{\mathcal{A}}(\mathcal{A}_2)$
\begin{align}\label{causalproperty}
S_{\mathcal{A}}(\mathcal{A}_2)=\mathcal{S}_{\mathcal{A}'}(\mathcal{A}'_2)\,.
\end{align}

\subsection{Testing the entanglement contour proposal}
It is proposed in \cite{Wen:2018whg} that in general theories the $s_{\mathcal{A}}(\mathcal{A}_2)$ can be written as a linear combination of the entanglement entropies of single subintervals inside $\mathcal{A}$
\begin{align}\label{proposal}
s_{\mathcal{A}}(\mathcal{A}_2)&=\frac{1}{2}\left( S_{\mathcal{A}_1\cup\mathcal{A}_2}+S_{\mathcal{A}_2\cup\mathcal{A}_3}-S_{\mathcal{A}_1}-S_{\mathcal{A}_3}\right)\,.
\end{align}
Here we would like to test this proposal for WCFT. Using \eqref{SeeLM} for all these sub-intervals on the straight interval $\mathcal{A}$ \eqref{straightA} we have
\begin{align}\label{I2S}
s_{\mathcal{A}}(\mathcal{A}_2)&=\frac{c}{6}\left(v'_2-v'_1+\frac{1}{2}\log\frac{(v'_2+\frac{l_v}{2})(v'_1-\frac{l_v}{2})}{(v'_2-\frac{l_v}{2})(v'_1+\frac{l_v}{2})}\right)=\frac{c}{6}(v_2-v_1)\,,
\end{align}
which coincide with the result \eqref{SA2} we get from the entanglement contour \eqref{contour}.

Then we test the causal property \eqref{causalproperty} for the proposal \eqref{proposal}. We let the two endpoints of $\mathcal{A}_2$ run along the boundary modular flow lines $\mathcal{L}_{v_1}$ and $\mathcal{L}_{v_2}$ that passing trough them (see Fig. \ref{A2}),
\begin{align}
(u'_1,v'_1)=(u'_1,v_1-\text{arctanh}\frac{2 u'_1}{l_u})\,,
\\
 (u'_2,v'_2)=(u'_2,v_2-\text{arctanh}\frac{2 u'_2}{l_u})\,,
\end{align}
The new subinterval $\mathcal{A}'_2$ passes through the same class of modular planes as $\mathcal{A}_2$, so satisfy \eqref{causalproperty}. With all the endpoints of subintervals known, we apply \eqref{SeeLM} and find
 \begin{align}
 S_{\mathcal{A}'_1\cup\mathcal{A}'_2}+S_{\mathcal{A}'_2\cup\mathcal{A}'_3}-S_{\mathcal{A}'_1}-S_{\mathcal{A}'_3}&=\frac{c}{3}\left(v_2-v_1\right)
 \cr
&= S_{\mathcal{A}_1\cup\mathcal{A}_2}+S_{\mathcal{A}_2\cup\mathcal{A}_3}-S_{\mathcal{A}_1}-S_{\mathcal{A}_3}\,.
\end{align}
This indicates that the linear combination in \eqref{proposal} reproduce the right causal property for the contour function.

\section{The saddle that connect the two null curves $\gamma_\pm$ }\label{appendixB}
Here we prove that the regulated curve $\mathcal{E}_{reg}$ is the saddle among all the geodesics that connect $\gamma_\pm$ \eqref{gammapm}. To prove this we need to calculate the proper distances between arbitrary two points in the bulk, then we fix the endpoints on $\gamma_+$ and $\gamma_-$ respectively and find out the saddle among all the geodesics. It is easier to start from calculating the proper length of arbitrary two points in Poincar\'{e} AdS, then we rewrite the distance in terms of the variables in the AdS space with nonzero temperatures via a coordinate transformation.

For simplicity we consider the Poincar\'{e} AdS$_3$ spacetime
\begin{align}\label{adsT0}
ds^2= \ell^2\Big(\frac{ d \rho^2}{4  \rho^2}+2  \rho d U d V\Big)\,.
\end{align}
with the geodesics given by
\begin{align}\label{geodesic}
U=&\frac{l_U}{2}\tanh \tau+c_U\,,
\qquad
 V=\frac{l_V}{2}\tanh \tau+c_V\,,
\qquad
\rho=\frac{2\cosh^2 \tau}{l_U l_V}\,.
\end{align}
Here $c_U$ and $c_V$ are arbitrary constants, while $l_U$ and $l_V$ are the distances between the endpoints on the boundary along the $U$ and $V$ directions respectively, $\tau$ is the parameter that parametrize the geodesic.  Along this line we have
\begin{align}
ds^2=\ell^2 d\tau^2
\end{align}
so the proper length is just
\begin{align}\label{tau1tau2}
L_{\text{AdS}}=\ell(\tau_1-\tau_2)
\end{align}
Note that any two spacelike separated points, for example $(U_1,V_1,\rho_1)$ and $(U_2,V_2,\rho_2)$, in the bulk can be connected by a geodesic line described by \eqref{geodesic}, thus the distance between them is just \eqref{tau1tau2}. Using (\ref{geodesic}), this proper length \eqref{tau1tau2} can be expressed in terms of the coordinates of the two endpoints
\begin{align}
&L_{\text{AdS}}\left(U_1,V_1,\rho_1,U_2,V_2,\rho_2\right)
\cr
=&\frac{1}{2} \log \left(\frac{\rho _2 \left(\rho _2+X\right)+\rho _1 \left(\rho _2 Y \left(2 \rho _2+X\right)+X\right)+\left(\rho _1+\rho _2 \rho _1 Y\right){}^2}{2 \rho _1 \rho _2}\right)
\end{align}
where
\begin{align}
Y=&2 \left(U_1-U_2\right) \left(V_1-V_2\right)
\cr
X=&\sqrt{\rho _1^2+2 \rho _2 \rho _1 \left(\rho _1 Y-1\right)+\left(\rho _2+\rho _1 \rho _2 Y\right){}^2}
\end{align}

Then we use the transformation from Poincar\'{e} AdS$_{(U,V,\rho)}$ to AdS$_{(u,v,r)}$ \eqref{btz} with nonzero temperatures $T_u$ and $T_v$
\begin{align}\label{ctgeneralads}
U=\,& e^{2 T_{u} u}\sqrt{1-\frac{2 T_{u} T_{v}}{r+T_{u} T_{v}}}\,,
\cr
V=\,& e^{2 T_{v} v}\sqrt{1-\frac{2 T_{u} T_{v}}{r+T_{u} T_{v}}}\,,
\cr
\rho=\,& \frac{(r+T_{u} T_{v})e^{-2(T_{u} u+T_{v} v)}}{4 T_{u} T_{v}}\,,
\end{align}
to re-express the distance in terms of the coordinates of the two endpoints in AdS$_{(u,v,r)}$
\begin{align}
L_{\text{AdS}}\left(u_1,v_1,r_1,u_2,v_2,r_2\right).
\end{align}
Then we set $T_u=0,~T_v=1$ (note that we should not set $T_u=0$ at first, or the first equation in \eqref{ctgeneralads} will be trivial) and set the endpoints on the null geodesics \eqref{gammapm}. Finally we find the distance $L_{\text{AdS}}\left(r_1,r_2\right)$ as a function of only $r_1$ and $r_2$.
We will not write down the explicit expression for $L_{\text{AdS}}\left(u_1,v_1,r_1,u_2,v_2,r_2\right)$ and $L_{\text{AdS}}\left(r_1,r_2\right)$ since they are very complicated. Solving the saddle points equation 
\begin{align}
\frac{\partial L_{\text{AdS}}\left(r_1,r_2\right)}{\partial r_1}=\frac{1-r_2 l_u}{\sqrt{\left(l_u \left(r_I^2 e^{2 l_v}-r_1 r_2\right)+r_1+r_2\right){}^2-4 r_I^2 e^{2 l_v}}}=0\,,
\cr
\frac{\partial L_{\text{AdS}}\left(r_1,r_2\right)}{\partial r_2}=\frac{1-r_1 l_u}{\sqrt{\left(l_u \left(r_I^2 e^{2 l_v}-r_1 r_2\right)+r_1+r_2\right){}^2-4 r_I^2 e^{2 l_v}}}=0\,,
\end{align}
we get
\begin{align}
r_1=r_2=\frac{1}{l_u}\,.
\end{align}
We see that the saddle is independent of $r_I$ and $l_v$. It is clear to see that the saddle geodesic is just our curve $\mathcal{E}_{reg}$.


\bibliographystyle{JHEP}
 \bibliography{lmbib}

\end{document}